\documentclass[twocolumn,twocolappendix]{aastex63}

\graphicspath{{./}{figures/}}
\usepackage{color}
\usepackage{soul}
\usepackage{natbib}
\usepackage{hyperref}
\usepackage{amsmath}
\usepackage{xspace}
\usepackage{graphicx}
\usepackage{enumitem}
\usepackage{amssymb}
\usepackage{xifthen}
\usepackage{hyperref}
\usepackage[normalem]{ulem}
\usepackage{comment}
\usepackage{float}
\usepackage{CJKutf8}
\usepackage{bbm}

\hypersetup{    
  colorlinks      = {true},
  linkcolor       = {blue},
  citecolor       = {blue},
  urlcolor        = {blue},
}

\newcommand{\code}[1]{\texttt{#1}\xspace}

\NewPageAfterKeywords

\makeatletter
\def\blfootnote{\xdef\@thefnmark{}\@footnotetext}
\makeatother

\usepackage{bm}
\usepackage{graphicx,pifont}
\let\oldding\ding
\renewcommand{\ding}[2][1]{\scalebox{#1}{\oldding{#2}}}

\usepackage{array}
\newcolumntype{H}{>{\setbox0=\hbox\bgroup}c<{\egroup}@{}}

\usepackage{natbib}
\defcitealias{Medina2025c}{M25}

\shorttitle{DESI DR2 Milky Way Mass with GME}

\shortauthors{Medina et al.}

\begin{document}

\title{A distribution-function estimate of the Milky Way mass profile from outer halo stars in DESI DR2 
}

\author[0000-0003-0105-9576]{Gustavo E. Medina}
\affiliation{Kapteyn Astronomical Institute, University of Groningen, P.O.Box 880, 9700 AV Groningen, The Netherlands}
\affiliation{Department of Astronomy \& Astrophysics, University of Toronto, Toronto, ON M5S 3H4, Canada}

\author[0000-0003-4952-1299, gname='Thomas', sname='Callingham']{T.~M.~Callingham}
\affiliation{Institute for Astronomy, University of Edinburgh, Royal Observatory, Blackford Hill, Edinburgh EH9 3HJ, UK}

\author[0000-0002-6257-2341, gname='Monica', sname='Valluri']{M.~Valluri}
\affiliation{Department of Astronomy, University of Michigan, Ann Arbor, MI 48109, USA}
\affiliation{University of Michigan, 500 S. State Street, Ann Arbor, MI 48109, USA}

\author[0000-0001-5805-5766]{Alexander~H.~Riley}
\affiliation{Lund Observatory, Division of Astrophysics, Department of Physics, Lund University, SE-221 00 Lund, Sweden}
\affiliation{Institute for Computational Cosmology, Department of Physics, Durham University, South Road, Durham DH1 3LE, UK}


\author[0000-0002-9110-6163, gname='Ting', sname='Li']{Ting~S.~Li}
\affiliation{Department of Astronomy \& Astrophysics, University of Toronto, Toronto, ON M5S 3H4, Canada}

\author{Jiaxin Han}
\affiliation{Department of Astronomy, School of Physics and Astronomy, Shanghai Jiao Tong University, Shanghai 200240, China}

\author[0000-0002-0740-1507, gname='Leandro', sname='Beraldo e Silva']{L.~{Beraldo e Silva}}
\affiliation{Observat\'orio Nacional, Rio de Janeiro - RJ, 20921-400, Brasil}

\author[0000-0001-9852-9954, gname='Oleg', sname='Gnedin']{Oleg~Y.~Gnedin}
\affiliation{Department of Astronomy, University of Michigan, Ann Arbor, MI 48109, USA}
\affiliation{University of Michigan, 500 S. State Street, Ann Arbor, MI 48109, USA}

\author[0000-0002-5689-8791, gname='Amanda', sname='Byström']{A.~Bystr\"om}
\affiliation{Institute for Astronomy, University of Edinburgh, Royal Observatory, Blackford Hill, Edinburgh EH9 3HJ, UK}

\author[0000-0003-2644-135X, gname='Sergey', sname='Koposov']{S.~E.~Koposov}
\affiliation{Institute for Astronomy, University of Edinburgh, Royal Observatory, Blackford Hill, Edinburgh EH9 3HJ, UK}
\affiliation{Institute of Astronomy, University of Cambridge, Madingley Road, Cambridge CB3 0HA, UK}

\author{A.~Slizewski}
\affiliation{Department of Astronomy and Astrophysics, University of Toronto, 50 St. George Street, Toronto, ON M5P 0A2, Canada}

\author{W.~Wang}
\affiliation{Department of Astronomy, School of Physics and Astronomy, Shanghai Jiao Tong University, Shanghai 200240, China}

\author[0000-0003-3734-8177, gname='Gwendolyn', sname='Eadie']{G.~M.~Eadie}
\affiliation{Department of Astronomy \& Astrophysics, University of Toronto, Toronto, ON M5S 3H4, Canada}
\affiliation{Department of Statistical Sciences, University of Toronto, Toronto, ON M5S 1Z5, Canada}

\author[0000-0002-7393-3595]{N.~R.~Sandford}
\affiliation{Department of Astronomy and Astrophysics, University of California, Santa Cruz, 1156 High Street, Santa Cruz, CA 95064, USA}
\affiliation{Department of Astronomy and Astrophysics, University of Toronto, 50 St. George Street, Toronto, ON M5P 0A2, Canada}

\author{B.~A.~Weaver}
\affiliation{NSF NOIRLab, 950 N. Cherry Ave., Tucson, AZ 85719, USA}

\author[0000-0002-7667-0081]{R.~G.~Carlberg}
\affiliation{Department of Astronomy \& Astrophysics, University of Toronto, Toronto, ON M5S 3H4, Canada}


\author{J.~Aguilar}
\affiliation{Lawrence Berkeley National Laboratory, 1 Cyclotron Road, Berkeley, CA 94720, USA}
\author[0000-0001-6098-7247]{S.~Ahlen}
\affiliation{Department of Physics, Boston University, 590 Commonwealth Avenue, Boston, MA 02215 USA}
\author[0000-0002-0084-572X]{C.~Allende~Prieto}
\affiliation{Departamento de Astrof\'{\i}sica, Universidad de La Laguna (ULL), E-38206, La Laguna, Tenerife, Spain}
\affiliation{Instituto de Astrof\'{\i}sica de Canarias, C/ V\'{\i}a L\'{a}ctea, s/n, E-38205 La Laguna, Tenerife, Spain}
\author[0000-0001-5998-3986]{A.~Aviles}
\affiliation{Instituto Avanzado de Cosmolog\'{\i}a A.~C., San Marcos 11 - Atenas 202. Magdalena Contreras. Ciudad de M\'{e}xico C.~P.~10720, M\'{e}xico}
\affiliation{Instituto de Ciencias F\'{\i}sicas, Universidad Nacional Aut\'onoma de M\'exico, Av. Universidad s/n, Cuernavaca, Morelos, C.~P.~62210, M\'exico}
\author[0000-0003-0467-5438]{F.~Beutler}
\affiliation{Institute for Astronomy, University of Edinburgh, Royal Observatory, Blackford Hill, Edinburgh EH9 3HJ, UK}
\author[0000-0001-9712-0006]{D.~Bianchi}
\affiliation{Dipartimento di Fisica ``Aldo Pontremoli'', Universit\`a degli Studi di Milano, Via Celoria 16, I-20133 Milano, Italy}
\affiliation{INAF-Osservatorio Astronomico di Brera, Via Brera 28, 20122 Milano, Italy}
\author{D.~Brooks}
\affiliation{Department of Physics \& Astronomy, University College London, Gower Street, London, WC1E 6BT, UK}
\author{R.~Canning}
\affiliation{Institute of Cosmology and Gravitation, University of Portsmouth, Dennis Sciama Building, Portsmouth, PO1 3FX, UK}
\author[0000-0003-3044-5150]{A.~Carnero Rosell}
\affiliation{Departamento de Astrof\'{\i}sica, Universidad de La Laguna (ULL), E-38206, La Laguna, Tenerife, Spain}
\affiliation{Instituto de Astrof\'{\i}sica de Canarias, C/ V\'{\i}a L\'{a}ctea, s/n, E-38205 La Laguna, Tenerife, Spain}
\author[0000-0001-7316-4573]{F.~J.~Castander}
\affiliation{Institut d'Estudis Espacials de Catalunya (IEEC), c/ Esteve Terradas 1, Edifici RDIT, Campus PMT-UPC, 08860 Castelldefels, Spain}
\affiliation{Institute of Space Sciences, ICE-CSIC, Campus UAB, Carrer de Can Magrans s/n, 08913 Bellaterra, Barcelona, Spain}
\author[0000-0001-8996-4874]{E.~Chaussidon}
\affiliation{Lawrence Berkeley National Laboratory, 1 Cyclotron Road, Berkeley, CA 94720, USA}
\author{T.~Claybaugh}
\affiliation{Lawrence Berkeley National Laboratory, 1 Cyclotron Road, Berkeley, CA 94720, USA}
\author[0000-0001-8274-158X]{A.~P.~Cooper}
\affiliation{Institute of Astronomy and Department of Physics, National Tsing Hua University, 101 Kuang-Fu Rd. Sec. 2, Hsinchu 30013, Taiwan}
\author[0000-0002-2169-0595]{A.~Cuceu}
\affiliation{Lawrence Berkeley National Laboratory, 1 Cyclotron Road, Berkeley, CA 94720, USA}
\author[0000-0002-1769-1640]{A.~de la Macorra}
\affiliation{Instituto de F\'{\i}sica, Universidad Nacional Aut\'{o}noma de M\'{e}xico,  Circuito de la Investigaci\'{o}n Cient\'{\i}fica, Ciudad Universitaria, Cd. de M\'{e}xico  C.~P.~04510,  M\'{e}xico}
\author[0000-0002-4928-4003]{Arjun~Dey}
\affiliation{NSF NOIRLab, 950 N. Cherry Ave., Tucson, AZ 85719, USA}
\author[0000-0002-5665-7912]{Biprateep~Dey}
\affiliation{Department of Astronomy \& Astrophysics, University of Toronto, Toronto, ON M5S 3H4, Canada}
\affiliation{Department of Physics \& Astronomy and Pittsburgh Particle Physics, Astrophysics, and Cosmology Center (PITT PACC), University of Pittsburgh, 3941 O'Hara Street, Pittsburgh, PA 15260, USA}
\author{P.~Doel}
\affiliation{Department of Physics \& Astronomy, University College London, Gower Street, London, WC1E 6BT, UK}
\author[0000-0003-4992-7854]{S.~Ferraro}
\affiliation{Lawrence Berkeley National Laboratory, 1 Cyclotron Road, Berkeley, CA 94720, USA}
\affiliation{University of California, Berkeley, 110 Sproul Hall \#5800 Berkeley, CA 94720, USA}
\author[0000-0002-2890-3725]{J.~E.~Forero-Romero}
\affiliation{Departamento de F\'isica, Universidad de los Andes, Cra. 1 No. 18A-10, Edificio Ip, CP 111711, Bogot\'a, Colombia}
\affiliation{Observatorio Astron\'omico, Universidad de los Andes, Cra. 1 No. 18A-10, Edificio H, CP 111711 Bogot\'a, Colombia}
\author[0000-0001-9632-0815]{E.~Gaztañaga}
\affiliation{Institut d'Estudis Espacials de Catalunya (IEEC), c/ Esteve Terradas 1, Edifici RDIT, Campus PMT-UPC, 08860 Castelldefels, Spain}
\affiliation{Institute of Cosmology and Gravitation, University of Portsmouth, Dennis Sciama Building, Portsmouth, PO1 3FX, UK}
\affiliation{Institute of Space Sciences, ICE-CSIC, Campus UAB, Carrer de Can Magrans s/n, 08913 Bellaterra, Barcelona, Spain}
\author[0000-0003-3142-233X]{Satya~{Gontcho A Gontcho}}
\affiliation{University of Virginia, Department of Astronomy, Charlottesville, VA 22904, USA}
\author{G.~Gutierrez}
\affiliation{Fermi National Accelerator Laboratory, PO Box 500, Batavia, IL 60510, USA}
\author[0000-0002-9136-9609]{H.~K.~Herrera-Alcantar}
\affiliation{Institut d'Astrophysique de Paris. 98 bis boulevard Arago. 75014 Paris, France}
\affiliation{Center for Cosmology and AstroParticle Physics, The Ohio State University, 191 West Woodruff Avenue, Columbus, OH 43210, USA}
\affiliation{Department of Astronomy, The Ohio State University, 4055 McPherson Laboratory, 140 W 18th Avenue, Columbus, OH 43210, USA}
\affiliation{IRFU, CEA, Universit\'{e} Paris-Saclay, F-91191 Gif-sur-Yvette, France}
\affiliation{The Ohio State University, Columbus, 43210 OH, USA}
\author[0000-0002-6024-466X]{M.~Ishak}
\affiliation{Department of Physics, The University of Texas at Dallas, 800 W. Campbell Rd., Richardson, TX 75080, USA}
\author[0000-0002-0000-2394]{S.~Juneau}
\affiliation{NSF NOIRLab, 950 N. Cherry Ave., Tucson, AZ 85719, USA}
\author{R.~Kehoe}
\affiliation{Department of Physics, Southern Methodist University, 3215 Daniel Avenue, Dallas, TX 75275, USA}
\author[0000-0002-8828-5463]{D.~Kirkby}
\affiliation{Department of Physics and Astronomy, University of California, Irvine, 92697, USA}
\author[0000-0001-6356-7424]{A.~Kremin}
\affiliation{Lawrence Berkeley National Laboratory, 1 Cyclotron Road, Berkeley, CA 94720, USA}
\author[0000-0002-1134-9035]{O.~Lahav}
\affiliation{Department of Physics \& Astronomy, University College London, Gower Street, London, WC1E 6BT, UK}
\author[0000-0002-6731-9329]{C.~Lamman}
\affiliation{The Ohio State University, Columbus, 43210 OH, USA}
\author[0000-0003-1838-8528]{M.~Landriau}
\affiliation{Lawrence Berkeley National Laboratory, 1 Cyclotron Road, Berkeley, CA 94720, USA}
\author[0000-0001-7178-8868]{L.~Le~Guillou}
\affiliation{Sorbonne Universit\'{e}, CNRS/IN2P3, Laboratoire de Physique Nucl\'{e}aire et de Hautes Energies (LPNHE), FR-75005 Paris, France}
\author[0000-0002-3677-3617]{A.~Leauthaud}
\affiliation{Department of Astronomy and Astrophysics, UCO/Lick Observatory, University of California, 1156 High Street, Santa Cruz, CA 95064, USA}
\affiliation{Department of Astronomy and Astrophysics, University of California, Santa Cruz, 1156 High Street, Santa Cruz, CA 95065, USA}
\author[0000-0002-1125-7384]{A.~Meisner}
\affiliation{NSF NOIRLab, 950 N. Cherry Ave., Tucson, AZ 85719, USA}
\author{R.~Miquel}
\affiliation{Instituci\'{o} Catalana de Recerca i Estudis Avan\c{c}ats, Passeig de Llu\'{\i}s Companys, 23, 08010 Barcelona, Spain}
\affiliation{Institut de F\'{i}sica d’Altes Energies (IFAE), The Barcelona Institute of Science and Technology, Edifici Cn, Campus UAB, 08193, Bellaterra (Barcelona), Spain}
\author[0000-0001-9070-3102]{S.~Nadathur}
\affiliation{Institute of Cosmology and Gravitation, University of Portsmouth, Dennis Sciama Building, Portsmouth, PO1 3FX, UK}
\author[0000-0002-4637-2868]{E.~Paillas}
\affiliation{Instituto de Estudios Astrof\'isicos, Facultad de Ingenier\'ia y Ciencias, Universidad Diego Portales, Av. Ej\'ercito Libertador 441, Santiago, Chile}
\author[0000-0003-3188-784X]{N.~Palanque-Delabrouille}
\affiliation{IRFU, CEA, Universit\'{e} Paris-Saclay, F-91191 Gif-sur-Yvette, France}
\affiliation{Lawrence Berkeley National Laboratory, 1 Cyclotron Road, Berkeley, CA 94720, USA}
\author[0000-0002-0644-5727]{W.~J.~Percival}
\affiliation{Department of Physics and Astronomy, University of Waterloo, 200 University Ave W, Waterloo, ON N2L 3G1, Canada}
\affiliation{Perimeter Institute for Theoretical Physics, 31 Caroline St. North, Waterloo, ON N2L 2Y5, Canada}
\affiliation{Waterloo Centre for Astrophysics, University of Waterloo, 200 University Ave W, Waterloo, ON N2L 3G1, Canada}
\author[0000-0001-7145-8674]{F.~Prada}
\affiliation{Instituto de Astrof\'{i}sica de Andaluc\'{i}a (CSIC), Glorieta de la Astronom\'{i}a, s/n, E-18008 Granada, Spain}
\author[0000-0001-6979-0125]{I.~P\'erez-R\`afols}
\affiliation{Departament de F\'isica, EEBE, Universitat Polit\`ecnica de Catalunya, c/Eduard Maristany 10, 08930 Barcelona, Spain}
\author[0000-0002-3500-6635]{C.~Ravoux}
\affiliation{Universit\'{e} Clermont-Auvergne, CNRS, LPCA, 63000 Clermont-Ferrand, France}
\author{G.~Rossi}
\affiliation{Department of Physics and Astronomy, Sejong University, 209 Neungdong-ro, Gwangjin-gu, Seoul 05006, Republic of Korea}
\author[0000-0002-9646-8198]{E.~Sanchez}
\affiliation{CIEMAT, Avenida Complutense 40, E-28040 Madrid, Spain}
\author[0000-0002-0408-5633]{C.~Saulder}
\affiliation{Max Planck Institute for Extraterrestrial Physics, Gie\ss enbachstra\ss e 1, 85748 Garching, Germany}
\author{D.~Schlegel}
\affiliation{Lawrence Berkeley National Laboratory, 1 Cyclotron Road, Berkeley, CA 94720, USA}
\author{M.~Schubnell}
\affiliation{Department of Physics, University of Michigan, 450 Church Street, Ann Arbor, MI 48109, USA}
\affiliation{University of Michigan, 500 S. State Street, Ann Arbor, MI 48109, USA}
\author[0000-0002-3461-0320]{J.~Silber}
\affiliation{Lawrence Berkeley National Laboratory, 1 Cyclotron Road, Berkeley, CA 94720, USA}
\author[0000-0002-2949-2155]{M.~Siudek}
\affiliation{Institute of Space Sciences, ICE-CSIC, Campus UAB, Carrer de Can Magrans s/n, 08913 Bellaterra, Barcelona, Spain}
\affiliation{Instituto de Astrof\'{\i}sica de Canarias, C/ V\'{\i}a L\'{a}ctea, s/n, E-38205 La Laguna, Tenerife, Spain}
\author[0000-0003-1704-0781]{G.~Tarl\'{e}}
\affiliation{University of Michigan, 500 S. State Street, Ann Arbor, MI 48109, USA}

\author[0000-0001-5381-4372]{R.~Zhou}
\affiliation{Lawrence Berkeley National Laboratory, 1 Cyclotron Road, Berkeley, CA 94720, USA}


\collaboration{100}{(DESI Collaboration)}


\correspondingauthor{Gustavo E. Medina}
\email{gustavo.medina@utoronto.ca}


\begin{abstract}
We present measurements of the Milky Way mass using blue horizontal-branch (BHB) and RR Lyrae (RRL) stars from the second data release of DESI (DR2), covering its first three years of operations. 
We employ 279 BHBs and 213 RRLs at $\sim\!50$--$80$\,kpc from the Galactic center, selected from catalogs roughly twice as large as those of DESI DR1, used in our previous mass measurement.
We apply the Galactic Mass Estimator (GME), a hierarchical Bayesian framework that assumes power-law forms for the potential and the tracer density, and accounts for distance and velocity uncertainties.
Validating the method on 24 AuriDESI mock catalogs (six Auriga halos at four viewing angles), we find that it overestimates the enclosed mass by $\sim\!15\%$ at 80\,kpc, and correct our results accordingly. 
We obtain $M(<80\,{\rm kpc}) = 0.64^{+0.06}_{-0.06}$ and $0.53^{+0.05}_{-0.05} \times 10^{12}\,M_\odot$ for BHBs and RRLs, respectively, while including Sagittarius stream stars lowers these masses by $\sim\!10$\%.
Extrapolating to 100\,kpc, we obtain $M(<100\,{\rm kpc}) = 0.74^{+0.07}_{-0.07}$ and $0.62^{+0.06}_{-0.07} \times 10^{12}\,M_\odot$, and anchoring an NFW halo to our GME constraints at 65\,kpc yields virial masses of $M_{200} = 1.12^{+0.14}_{-0.12}$ and $0.89^{+0.11}_{-0.10} \times 10^{12}\,M_\odot$.
BHBs yield systematically higher masses than RRLs across the full radial range, reflecting the impact of tracer choice even within a single method.
This work is part of a wider DESI DR2 Key Project effort to measure the Milky Way mass and constrain systematics arising from different samples and methodologies.
\end{abstract}






\section{Introduction} \label{sec:intro}

The total mass of the Milky Way is a fundamental ingredient to understand its hierarchical assembly and its place in the universe (e.g., from comparisons with cosmological simulations; \citealt{Bland-Hawthorn2016, Callingham2019,Wang2020}).
Despite its importance, the total Milky Way mass remains uncertain by nearly a factor of two, with published estimates spanning $\sim\!0.5$--$2.0 \times 10^{12}\,M_{\odot}$, depending on the methodology and the tracer population used for the mass derivation and the radial range probed \citep[e.g.,][]{Bland-Hawthorn2016, Watkins2019, Deason2021,Hunt2025}.  

A variety of kinematic methods have been employed to constrain the Milky Way mass out to large radii, including the escape velocity technique \citep[e.g.,][]{Piffl2014, Deason2019,Prudil2022}, Jeans equation modeling \citep[e.g.,][]{Xue2008,Gnedin2010}, and phase-space distribution function (DF) fitting \citep[e.g.,][]{Gibbons2014, Watkins2019, Bird2019}, including applications to satellite galaxies \citep{Li2019_mass, Li2020} and globular clusters \citep{Li2025}.
All of these methods rely on discrete tracers with measured distances and kinematics (proper motions and line-of-sight velocities), among which halo stars, globular clusters, or satellite galaxies are typically used. 

One of the main limitations of the use of tracers in the outer halo 
(at Galactocentric radii $R_{\rm GC} \gtrsim 20$\,kpc) is the scarcity of tracers with complete and precise phase-space information, since distance and proper motion measurements at such distances remain challenging even with state-of-the-art space missions and telescopes (e.g., {\it Gaia}).
Furthermore, line-of-sight velocity measurements of statistically significant samples often rely on large-scale spectroscopic surveys mapping the halo.   
Beyond observational limitations, the interpretation of outer-halo kinematics is further complicated by physical effects that violate the assumptions of equilibrium and spherical symmetry underlying most mass estimators, most notably the gravitational influence of the Large Magellanic Cloud \citep[e.g.,][]{Garavito-Camargo2019, Petersen2021} and the presence of tidal substructure and stellar streams \citep[e.g.,][]{Helmi1999, Wang2015,Wang2017,Wang2018,Han2016,Ibata2021}.

Among DF-based approaches, the Galactic Mass Estimator \citep[GME;][]{Eadie2015, Eadie2017} is a hierarchical Bayesian framework that naturally handles missing proper motions and heterogeneous measurement uncertainties by treating unknown tangential velocity components as nuisance parameters in the model.
Applied to globular clusters observed with {\it Gaia} DR2, it yielded an enclosed mass within $\sim\!125$\,kpc of $0.52^{+0.14}_{-0.10} \times 10^{12}\,M_{\odot}$ \citep{Eadie2019}.
The framework has since been extended to dwarf galaxies \citep{Slizewski2022} and to halo field stars \citep{Shen2022}, the latter supported by a publicly available Python implementation employing Hamiltonian Monte Carlo sampling and applied to K~giants from the H3 survey and {\it Gaia}~EDR3, yielding $M_{200} = 1.08^{+0.12}_{-0.11} \times 10^{12}\,M_{\odot}$\footnote{$M_{200}$ is defined as the mass enclosed within the radius at which the average density of the halo equals 200 times the critical density of the universe \citep[see e.g.,][]{Navarro1997,Bland-Hawthorn2016}.}. 
Most recently, \citet{Medina2025c} applied this method to 330 
BHBs and 110 RRLs from 
the first data release of the 
Dark Energy Spectroscopic Instrument \citep[DESI DR1;][]{DESI2016a, DESI2016b, Cooper2023}, 
covering $R_{\rm GC} \sim 50$--$100$\,kpc and obtaining enclosed masses within 100\,kpc of $0.57^{+0.08}_{-0.07} \times 10^{12}\,M_{\odot}$ and $0.55^{+0.12}_{-0.10} \times 10^{12}\,M_{\odot}$ for BHBs and RRLs, respectively.

The DESI survey 
is ideally suited to Milky Way mass studies, delivering spectroscopic observations of millions of halo stars over a wide sky area and providing full 6D phase-space information (i.e., 3D positions and 3D velocities) when combined with {\it Gaia} proper motions. 
Among the stellar tracers accessible to DESI, RR Lyrae stars (RRLs) and blue horizontal-branch stars (BHBs) are particularly valuable, as well-calibrated standard candles that yield precise heliocentric distances (with uncertainties at the $\sim\! 5\%$ and $\sim\!10\%$ level, respectively; see e.g., \citealt{Garofalo2022} and \citealt{Deason2011b}).
For the former, DESI's multi-epoch observations enable robust systemic velocity estimates despite the intrinsic pulsational variability of these stars \citep[see][]{Cooper2023, Koposov2024, Medina2025a}.  

This work is the direct continuation of \citet{Medina2025c} (hereafter \citetalias{Medina2025c}), which first applied the GME method to RRL and BHB outer-halo tracers in DESI DR1.
Here we extend that analysis to DESI's second data release (DR2), which encompasses the first three years of survey operations, with a significantly larger sample of tracers extending to $\sim\!120$\,kpc over a larger and more homogeneous on-sky footprint. 
This paper presents one of three independent Milky Way mass measurements carried out as part of the DESI Year 3 Key Project (Y3KP).
These complementary analyses use independent methods on the same tracer populations, and are presented by \citet{Valluri:2026nimble} and \citet{Zhou:2026empdf}. 
Our combined results are discussed by \citet{DESI:2026mwmass}.

This paper is organized as follows.
In Section~\ref{sec:data}, we describe the DESI DR2 BHB and RRL samples used in this work, including distance determinations, radial velocity measurements, and the selection cuts applied to the catalogs. Section~\ref{sec:model} summarizes the GME framework and the hierarchical Bayesian model adopted in our analysis. We validate the method on cosmological simulations in Section~\ref{sec:mocks}, using mock BHB and RRL catalogs from the AuriDESI simulation suite, 
spanning a representative range of halo assembly histories and viewing angles. The mass estimates derived from the DESI DR2 samples are presented in Section~\ref{sec:desiY3-models}. 
In Section~\ref{sec:discussion}, we examine the sensitivity of the mass recovery to selection cuts and other potential biases, compare our results with the DESI DR1 measurements of \citetalias{Medina2025c}, discuss the consistency between the BHB and RRL mass profiles, and place our estimates in the context of the broader literature. 
We summarize our findings in Section~\ref{sec:conclusions}.

\begin{figure*}
        \centering    
        \includegraphics[width=0.49\textwidth]{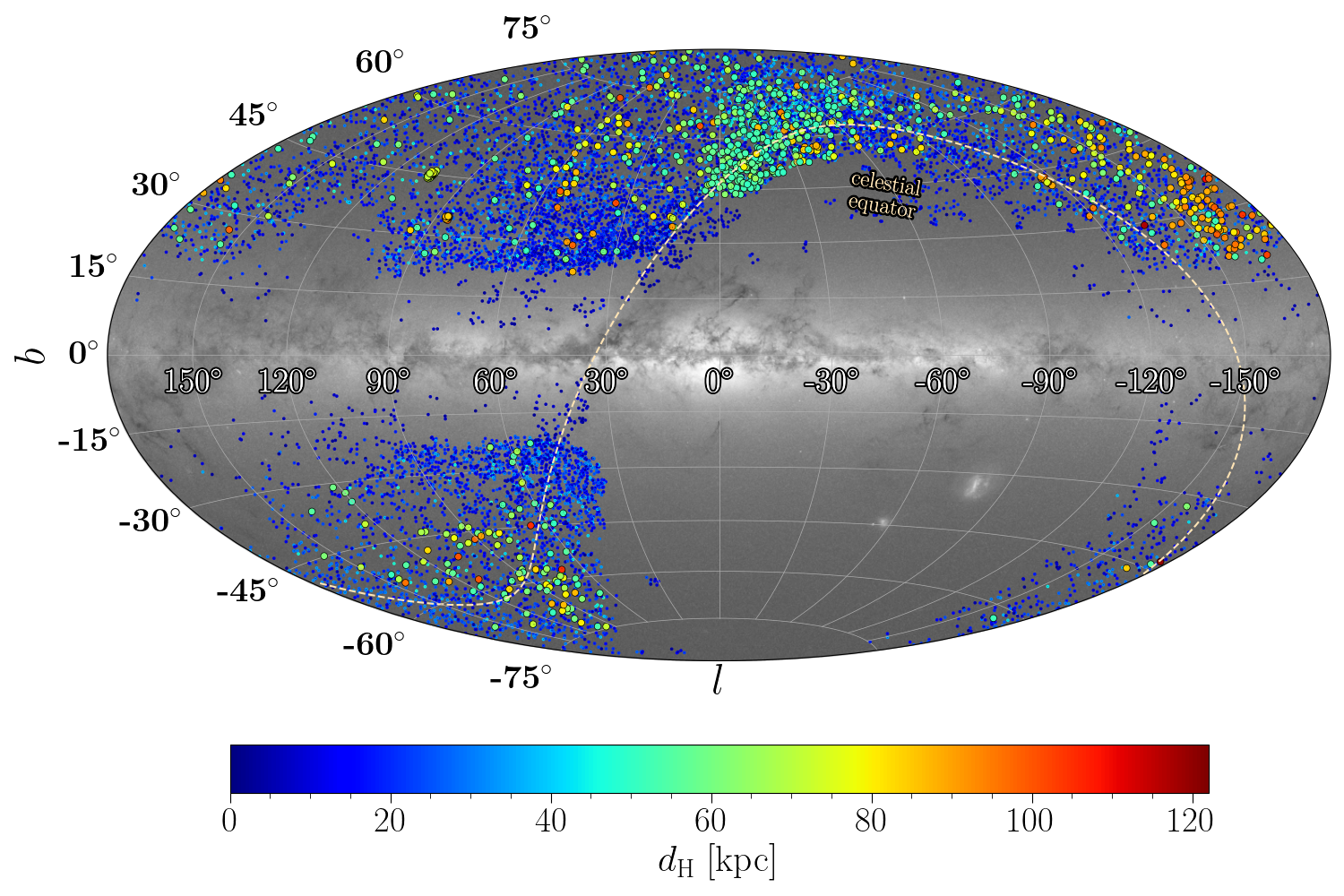}
        \includegraphics[width=0.49\textwidth]{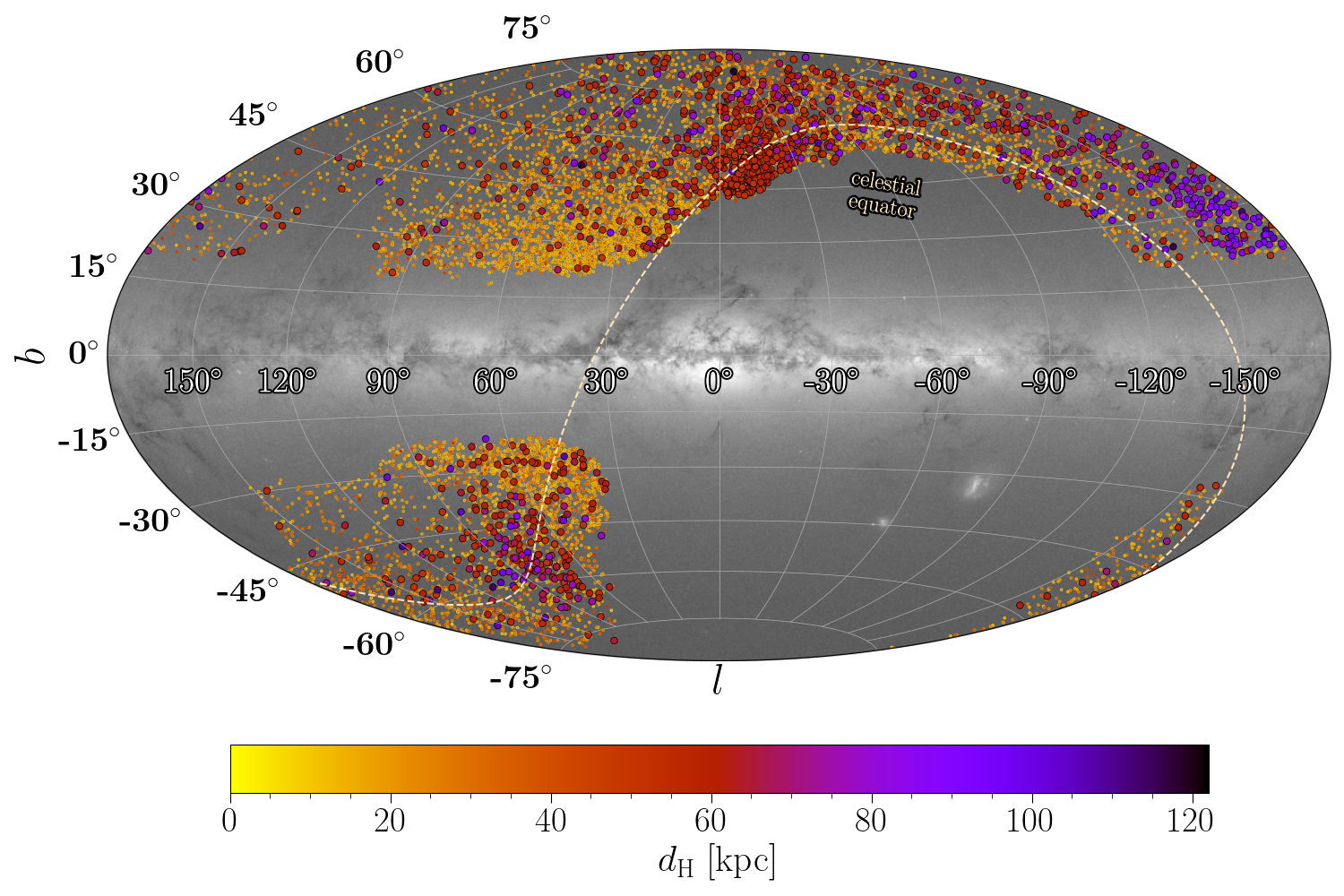}
    \caption{Spatial distribution of the full DESI Y3 RRL ({\it Left}) and BHB samples ({\it Right}), color-coded by heliocentric distance. 
    In both panels, the {\it Gaia} all-sky map is shown in the background as a reference. 
    \textit{Background image credit: Gaia Data Processing and Analysis Consortium (DPAC); A. Moitinho / A. F. Silva / M. Barros / C. Barata, University of Lisbon, Portugal; H. Savietto, Fork Research, Portugal.} 
    }
    \label{fig:map_RRLs_BHBs}
\end{figure*}

\section{Data}
\label{sec:data}

\subsection{The DESI survey}

The Dark Energy Spectroscopic Instrument \citep[DESI;][]{DESI2016a, DESI2016b, DESI2022} is a massively multiplexed fiber-fed spectrograph mounted at the prime focus of the Mayall 4\,m telescope at Kitt Peak National Observatory. 
It deploys 5,000 robotic fibers that feed 10 thermally controlled three-channel spectrographs \citep{DESI_Fiber_System_2024, DESI_Corrector_2024}, covering a wavelength range of 3,600--9,800\,\AA\ at spectral resolutions $R \sim 2{,}500$--$5{,}000$. 
Spectra are reduced and wavelength-calibrated using the dedicated DESI spectroscopic pipeline \citep{Guy2023,DESI_Survey_Operation_2023}.

Although primarily designed to study large-scale structure and the early phases of the Universe's evolution \citep[see e.g.,][]{DESI_DR1_cosmology_2024,DESI_DR2_cosmology_2025}, DESI also includes a dedicated Milky Way Survey \citep[MWS;][]{Cooper2023},
targeting millions of thick disk and halo stars at Galactic latitudes $|b| > 20\degr$.  
DESI observations are organized into surveys and programs.
The DESI surveys encompass the commissioning phase (CMX), three survey-validation stages (SV), and the 5-year main survey, and its programs, defined as \textsc{dark}, \textsc{bright}, and \textsc{backup}, correspond to the observing conditions in which each exposure is carried out \citep{Guy2023}. The DESI MWS is primarily conducted during \textsc{bright} time, when observing faint extragalactic targets is inefficient \citep{Cooper2023}.
Two dedicated spectral analysis pipelines, namely RVS (based on the {\sc rvspec} spectral analysis tool; \citealt{Koposov2019}) and SP (based on {\sc FERRE}; see e.g., \citealt{AllendePrieto2006}) are employed to derive radial velocities, effective temperatures ($T_{\rm eff}$), surface gravities ($\log g$), and iron abundances ([Fe/H]) for each observed star. 
In this work, we employ the results derived with the RVS pipeline. 
Early MWS data products based on RVS are presented by \citet{Koposov2024} and \citet{Koposov2025}.

In this work, we use data from DESI DR2, 
encompassing the first three years of survey operations and substantially expanding the sky coverage, number of observed targets, and number of repeat observations relative to DR1 \citep{DESI_DR1_2025}, which was used in the precursor work to this study, \citetalias{Medina2025c}.

\subsection{The RRL and BHB samples}
\label{sec:RRL-BHB-samples}

The DESI DR2 RRL catalog is built from a crossmatch between DESI and the {\it Gaia} DR3 RRL catalog \citep[][]{Clementini2023}.
It contains 12,868 stars, observed over a total of 28,410 individual epochs. 
This represents a factor of two improvement in the number of observed RRLs with respect to DESI DR1 \citep{Medina2025a,Medina2025b}.
Of these, 5,790 stars were observed only once, 3,345 have two epochs, 1,688 have three epochs, and 1,863 have more than three epochs. 
A full description of the catalog, including the target selection, the spectroscopic analysis of the sample, and their radial velocity variation correction, will be presented in forthcoming papers (Medina et al., in prep. and Section 2.3 in 
\citealt{DESI:2026mwmass}).

The DESI DR2 BHB catalog contains 12,989 stars in total. 
For stars observed more than once, only the best individual observation is retained (as identified by the \texttt{PRIMARY=True} flag in the RVS pipeline), resulting in an initial working sample of 10,445 BHBs. 
For a detailed description of the selection of BHBs in DESI, we refer the reader to \citet{Bystroem2025} and Section 2.2 in \citet{DESI:2026mwmass}.

Figure~\ref{fig:map_RRLs_BHBs} shows the spatial distribution of the full RRL and BHB samples in Galactic coordinates ($l, b$), color-coded by heliocentric distance (whose determination is presented in Section~\ref{sec:distances}). 
The Sagittarius (Sgr) stream is clearly visible in both panels as an overdensity of stars tracing a great circle across the sky, particularly near the celestial equator. 

\subsection{Distance determination}
\label{sec:distances}

The heliocentric distances ($d_{\rm H}$) of the BHB and RRL in our sample are obtained via distance modulus, using magnitude-metallicity  and absolute magnitude-color relations available in the literature. 

For RRLs, absolute magnitudes are computed using the absolute magnitude-metallicity relation from \citet{Garofalo2022} and intensity-averaged $G$ magnitudes ({\sc int\_average\_g} in the {\it Gaia} catalog), corrected by extinction. 
For this relation, we employ the iron abundances estimated by DESI's RVS pipeline per star, averaged over the available individual epochs.

For BHBs, we adopt the absolute magnitude-color relations adopted by \citet{Bystroem2025}, which rely on extinction-corrected $g$ and $r$ magnitudes in the Dark Energy Camera (DECam) photometric system from the DESI Legacy Imaging Survey \citep{2019AJ....157..168D}.

We utilize the dust maps of \citet{Schlafly2011} and assume the standard value relative visibility in the diffuse interstellar medium, $R_V = 3.1$ \citep{Schultz1975,Cardelli1989}, along with $A_G/A_V = 0.85926$\footnote{\href{http://stev.oapd.inaf.it/cgi-bin/cmd_3.3}{http://stev.oapd.inaf.it/cgi-bin/cmd\_3.3}}.
Figure~\ref{fig:map_RRLs_BHBs} shows the spatial distribution of our BHB and RRL samples color-coded by $d_{\rm H}$, which extend to approximately $120$\,kpc for both tracers.
We note that the distance uncertainties of our samples are smaller than 5\% for RRLs (for $\sim$99\% of the sample, from error propagation) and of $\sim6$\% for BHBs (following the distance validation of \citealt{Bystroem2025}).

\subsection{Radial velocities}
\label{sec:RVs}

The variable (pulsating) nature of RRLs makes the analysis of their kinematics challenging, since their rapidly changing atmospheres can reach $>100$\,$\mathrm{km\,s^{-1}}$ variations throughout their pulsation cycle \citep[see e.g.,][]{Sesar2012,Braga2021,Prudil2024b}. 
To mitigate this effect, the observed line-of-sight velocity of the RRLs in our sample is corrected for their pulsation component. 
The method adopted relies on \citetalias{Medina2025c}'s Bayesian approach, which uses single-epoch RVS velocities to construct radial velocity curve models and enables robust systemic velocity estimates even from individual observations (as long as their pulsation phase is available). 
As done by \citetalias{Medina2025c}, we rely on the ephemeris of each star as listed in the {\it Gaia} DR3 RRL catalog and DESI's mid-exposure observation times.

For our BHB sample, we adopt the velocities derived by the RVS pipeline from their coadded spectra as line-of-sight velocities.

\subsection{Completeness and selection cuts}
\label{sec:cuts}

Following \citet{Medina2025c}, we restrict the RRL sample to stars with intensity-averaged magnitudes {\sc int\_average\_g}~$<20$, which corresponds to $d_{\rm
H}\sim80$\,kpc for a typical outer-halo RRL ([Fe/H]~$=-2.0$, 
$E(B-V)<0.05$). 
Beyond this limit the completeness of the {\it Gaia} RRL catalog, on which the DESI RRL selection is based, drops below $\sim$80\%.
\citet{Mateu2020} find median completeness values for {\it Gaia} DR2 RRab (RRc) stars at $|b|>20^\circ$ of 92 (80)\% at $G=19.5$
($\sim$65\,kpc), decreasing to 85 (67)\% at $G=20$ ($\sim$80\,kpc), and \citet{Mateu2024} find that the completeness of the {\it Gaia} DR3 catalog falls below 80 (70)\% for RRab (RRc) beyond $\sim$60--80\,kpc. This is consistent with the 80\% completeness at 80\,kpc reported by \citet{Sesar2017} for the PS1 RRL catalog.

For both RRLs and BHBs, we only consider stars observed as part of DESI's main survey in the \textsc{bright} program. 
We note that this differs from \citetalias{Medina2025c}, who included stars observed in all DESI programs (including dark-time and survey-validation observations). 
Restricting the sample to this program roughly halves the number of BHBs and RRLs beyond 50\,kpc. 
As stated by \citet{Bystroem2025}, the BHB catalog has uniform completeness as a function of magnitude and color when using this criterion.
For RRLs, the $G<20$ limit adopted above also roughly corresponds to the magnitude limit of the DESI RRLs in the main \textsc{bright} program,  
and we therefore assume that, within this limit, our RRL sample is reasonably complete as a function of magnitude.

We base our selection cuts on those  
adopted by \citetalias{Medina2025c}, to remain consistent with the assumptions of our model (see Section~\ref{sec:model}). 
In brief, we retain stars with Galactocentric distances $R_{\rm GC} > 50$\,kpc and mean signal-to-noise ratio (across the three arms of the spectrograph) $\mathrm{SNR} > 3$.
We further discard known members of dwarf galaxies and globular clusters, based on a 0.5\,arcsec on-sky crossmatch with the members reported by \citet{vasiliev_GCs} and \citet{Pace2022}, and remove candidate members of the Sgr stream identified on the basis of their positions and distances (following \citetalias{Medina2025c} but without accounting for line-of-sight velocities). 
For the BHB sample we additionally require no warnings from the RVSpecFit pipeline, i.e., RVS\_WARN$=0$ \citep{rvspecfit,Koposov2025}.
Of the 352 BHB candidates that pass our other cuts, 73 (21\%) have RVS\_WARN~$\neq0$, 63 of them because the fitted projected rotational velocity is large.
These 73 stars also have lower S/N, larger radial-velocity uncertainties, 
and higher metallicities than the retained ones.
Since such contaminants are intrinsically fainter than BHB stars, their distances, and hence their Galactocentric radii and tangential velocities, would be overestimated, which would bias the inferred mass high.
Thus, this selection cut yields similar results as those from the velocity cuts applied by \citetalias{Medina2025c}, which were justified as requirements given the single power-law potential assumption of GME (i.e., unreliable in the inner halo) and the assumption of boundness of the tracers.

We only consider RRLs with $5000<T_{\rm eff}\ {\rm [K]}<8500$ and $0<\log g\ {\rm [dex]}<3.8$ (where $T_{\rm eff}$ and $\log g$ are the values reported by the RVS pipeline), as stars outside of these limits appear as clear outliers in Kiel space and are likely contaminants in our sample. 
Two additional RRLs, with {\it Gaia} DR3 source\_ids 2708106729874816896 and 382100910792267136, are also removed from the sample, as they appear as outliers with respect to the rest of the RRL $T_{\rm eff}$--$\log g$ distribution. 
For BHBs, we require colors $-0.3 \leq (g-r) \leq 0.0$ and stellar parameters in the ranges $7{,}000 < T_{\rm eff}/{\rm K} < 14{,}500$ and $1.50 < \log g - 0.00014\,T_{\rm eff} < 2.65$, following the classification criteria of \citet{Bystroem2025}. 

\begin{figure}
        \centering    
        \includegraphics[width=0.49\textwidth]{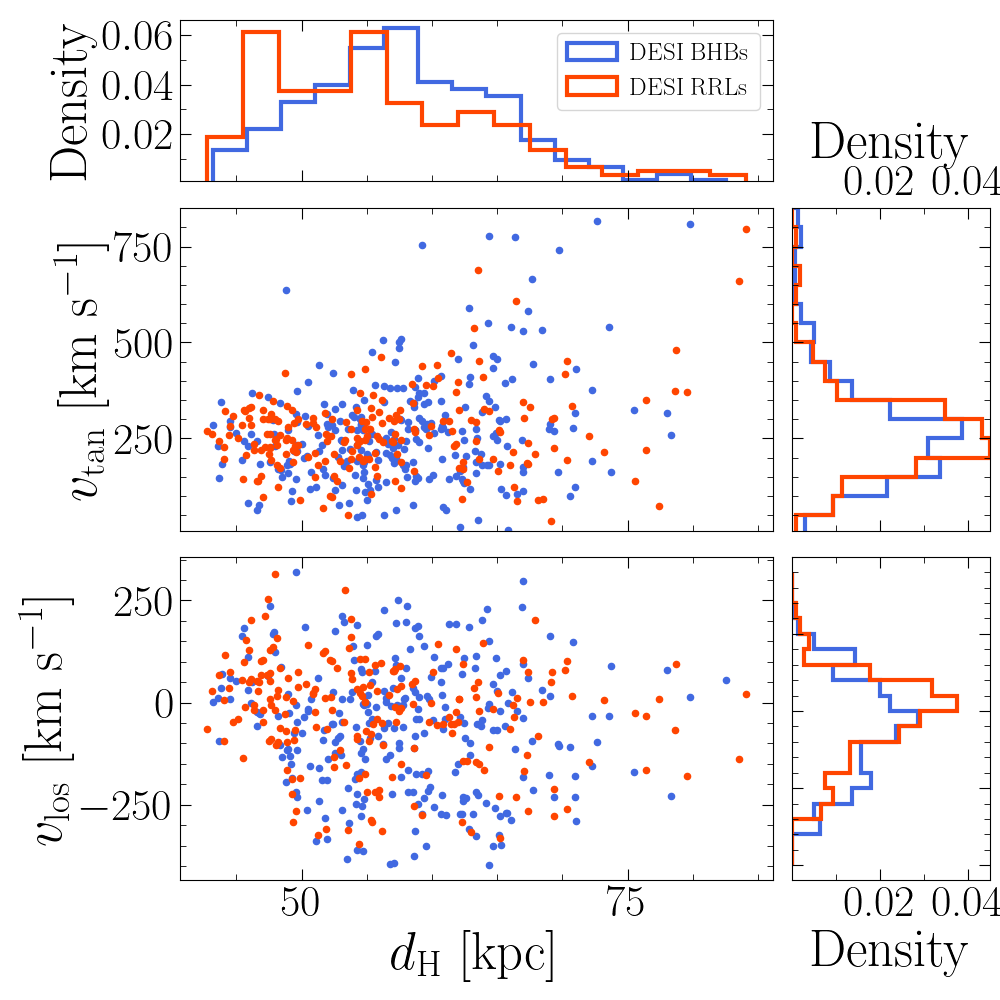}
    \caption{
    Distribution of heliocentric distances ($d_{\rm H}$), 
    tangential velocities ($v_{\rm tan}$), 
    and line-of-sight velocities ($v_{\rm los}$) of the DESI DR2 BHBs, after the selection cuts described in Section~\ref{sec:cuts}.
    Most BHBs with $v_{\rm tan}>500\,{\rm km\,s^{-1}}$ are removed by our RVS\_WARN cut. The remaining ones are mostly consistent with lower velocities within uncertainties (which are dominated by their proper motion uncertainties).
    }
    \label{fig:dist_and_velocity_distribution}
\end{figure}

The aforementioned selection cuts result in a sample of 213 RRLs and 279 BHBs, which we use as inputs for our model (Section~\ref{sec:desiY3-models}).
The heliocentric distance, line-of-sight velocity, and tangential velocity of our final samples are displayed in Figure~\ref{fig:dist_and_velocity_distribution}.

\section{The model} 
\label{sec:model}

Our model follows \citetalias{Medina2025c} 
and employs the Galactic Mass Estimator (GME) algorithm originally developed by \citet{Eadie2015, Eadie2017} and subsequently extended by \citet{Shen2022}. 
We summarize the key components below and refer the reader to \citetalias{Medina2025c} and the original references for a full description of the method.
   
The Galaxy is modeled as a spherically symmetric system in dynamical equilibrium, with a power-law gravitational potential:

\begin{equation}
  \Phi(r) = \Phi_0\, r^{-\gamma},
  \label{eq:potential} 
\end{equation}

\noindent where $\Phi_0$ is a scale factor and $\gamma$ is the potential slope, both treated as free parameters. 
The tracer population is assumed to follow a power-law radial density profile \citep{Little1987, Evans1997}:

\begin{equation}
  \rho(r) \propto r^{-\alpha},
  \label{eq:density}   
\end{equation}

\noindent where $\alpha$ is the density slope. 
As discussed by \citet{Shen2022} and \citetalias{Medina2025c}, $\alpha$ does not necessarily reflect the true physical distribution of the tracers in the standard GME method and is treated primarily as a nuisance parameter in their works. 
We emphasize that these single power-law forms are only assumed to hold over the radial range spanned by our tracers after selection cuts, i.e., $50 \lesssim R_{\rm GC} \lesssim 80$\,kpc (see Section~\ref{sec:cuts}), and are not intended to describe the Galaxy at smaller or larger radii.
Thus, enclosed masses quoted beyond this range (e.g., at 100\,kpc or $M_{200}$) are extrapolations of the fitted profiles (see Section~\ref{sec:desiY3-models}).

Given the potential in Equation~\ref{eq:potential}, the mass enclosed within a Galactocentric radius $r$ is:

\begin{multline}
  M(<r) = 2.325 \times 10^{-3}\, \gamma\,  
  \left(\frac{\Phi_0}{10^4\,\mathrm{km^2\,s^{-2}\,kpc^{\gamma}}}\right) \\
  \times \left(\frac{r}{\mathrm{kpc}}\right)^{1-\gamma}
  \times 10^{12}\,M_{\odot}.
  \label{eq:enclosedmass}  
\end{multline}  

Since $M(<r)$ depends on both $\Phi_0$ and $\gamma$, the enclosed mass within $50 \lesssim R_{\rm GC} \lesssim 80$\,kpc is mainly constrained by a combination of the two parameters.

The phase-space distribution function (DF) is expressed as a function of the  relative energy $\mathcal{E}$ and total angular momentum $L$ \citep{Little1987, Evans1997, Binney2008}:   


\begin{multline}
	\mathcal{F}(\mathcal{E}, L;\, \alpha, \beta, \gamma, \Phi_0) \;=\;
	\frac{
	L^{-2\beta}\,
	\mathcal{E}^{\,\beta(\gamma - 2)/\gamma\, +\, \alpha/\gamma\, -\, 3/2}
	}{
	\sqrt{8\pi^3}\; 2^{-2\beta}\; \Phi_0^{\,\alpha/\gamma\, -\, 2\beta/\gamma}
	} \\
	\times \frac{
	\Gamma\!\left(\alpha/\gamma - 2\beta/\gamma + 1\right)
	}{
	\Gamma\!\left(\beta(\gamma-2)/\gamma + \alpha/\gamma - 1/2\right)\,
	\Gamma(1 - \beta)
	},
\label{eq:df}
\end{multline}

\noindent where $\Gamma$ is the gamma function and $\beta$ is the velocity anisotropy parameter \citep{Binney2008}:  

\begin{equation} 
  \beta \equiv 1 - \frac{\sigma_\theta^2 + \sigma_\phi^2}{2\,\sigma_r^2},
  \label{eq:beta}  
\end{equation}

\noindent with $\sigma_r$, $\sigma_\theta$, and $\sigma_\phi$ represent the velocity dispersions in the radial, polar, and azimuthal directions, respectively. 
Based on this definition, isotropic orbits correspond to $\beta = 0$, while radially dominated orbits exhibit $\beta > 0$, and tangentially dominated orbits $\beta < 0$. 
In our model, $\beta$ is assumed to be constant across the radial range of the tracers used. 
The model is therefore fully characterized by four parameters: $\boldsymbol{\theta} = (\Phi_0,\, \gamma,\, \alpha,\, \beta)$.  
   
In the standard GME formulation, the DF is normalized implicitly over all space, without accounting for the minimum Galactocentric radius ($r_{\rm min}$) of the tracer sample. 
Since our sample is restricted to $R_{\rm GC} \geq r_{\rm min} = 50$\,kpc, this leads to a normalization that implicitly extends over a volume beyond the boundaries of the data used. 
This, in turn, translates into systematically biased posteriors for $\alpha$ that appear truncated at the lower boundary of the prior (as reported by \citealt{Shen2022} and \citetalias{Medina2025c}).
To account for the actual survey boundary, we introduce an additional term in the denominator of the DF 
that accounts for the spatial volume covered by the tracer samples. 
For a power-law density $\rho \propto r^{-\alpha}$, the expected number of tracers in the observed volume is proportional to the spatial Poisson normalization integral:
  
\begin{equation}   
\mathcal{N}(r_{\rm min}) = 4\pi \int_{r_{\rm min}}^{\infty}
r^{2-\alpha}\, dr = \frac{4\pi\, r_{\rm min}^{3-\alpha}}{\alpha - 3},  
\quad \alpha > 3,  
\label{eq:poisson_norm}
\end{equation} 

Dividing the DF by $\mathcal{N}(r_{\rm min})$ renormalizes the probability distribution to the survey volume and, as a consequence, makes $\alpha$ physically interpretable as the actual density slope of the outer-halo tracer population for $R_{\rm GC} > r_{\rm min}$, rather than a pure nuisance parameter. 
We note that, in practice, the volume probed by each tracer is also limited at large distances by the magnitude limits imposed on our samples.
Since the AuriDESI mocks used to validate our method (Section~\ref{sec:mocks}) are subject to the same magnitude limits, the effect of this finite survey depth on the recovered masses is accounted for, to first order, by the mock-based correction described in Section~\ref{sec:mockmass}.

In this work, we apply this adapted model to the DESI DR2 BHB and RRL samples.
Additional updates, including a treatment of the survey selection function, a scheme to identify and exclude potentially unbound tracers, and a correction for the gravitational influence of the Large Magellanic Cloud, will be implemented and validated in a forthcoming companion paper (Slizewski et al., in prep.).

\subsection{Hierarchical Bayesian Model}
\label{sec:hierarchicalbayesianmodel}

As in \citetalias{Medina2025c}, we adopt a hierarchical Bayesian framework that propagates observational uncertainties in the heliocentric distances and line-of-sight velocities of individual stars into the inference of the global model parameters \citep{Eadie2017, Shen2022}. 
The full posterior distribution over $\boldsymbol{\theta}$ given the measured distances $\mathbf{d}_{\rm H}$, line-of-sight velocities $\mathbf{v}_{\rm los}$, and astrometry
  $\mathbf{y}$, and their measurement uncertainties $\boldsymbol{\Delta}$, is:

\begin{multline}
	p(\boldsymbol{\theta} \mid \mathbf{d}_{\rm H}, \mathbf{v}_{\rm los}, \mathbf{y}, \boldsymbol{\Delta}) \propto
	p(\alpha)\, p(\beta)\, p(\Phi_0, \gamma) \\
	\times \prod_{i=1}^{N}
	p(d_{{\rm H},i} \mid d^\dagger_{{\rm H},i},\, \Delta d_{{\rm H},i})\,
	p(v_{{\rm los},i} \mid v^\dagger_{{\rm los},i},\, \Delta v_{{\rm los},i}) \\
	\times\, p(\mathbf{y}_i \mid \boldsymbol{\vartheta}_i,\, \boldsymbol{\Sigma}_i)\,
	p\big(h(d^\dagger_{{\rm H},i}, v^\dagger_{{\rm los},i}, \boldsymbol{\vartheta}_i) \mid \boldsymbol{\theta}\big),
	\label{eq:posterior}
\end{multline}

\noindent where $d_{{\rm H},i}$ and $v_{{\rm los},i}$ are the measured heliocentric distance and line-of-sight velocity of star $i$, with uncertainties $\Delta d_{{\rm H},i}$ and $\Delta v_{{\rm los},i}$, and $d^\dagger_{{\rm H},i}$ and $v^\dagger_{{\rm los},i}$ are their true (unknown) values.
Similarly, $\mathbf{y}_i = (\alpha_{*,i}, \delta_i, \mu_{\alpha*,i}\cos\delta_i, \mu_{\delta,i})$ are the measured right ascension, declination, and proper motion components\footnote{As in \citetalias{Medina2025c}, we use $\alpha_*$ for right ascension in this notation to avoid confusion with the density slope $\alpha$.}, $\boldsymbol{\vartheta}_i$ their true values, and $\boldsymbol{\Sigma}_i$ their covariance matrix.
All measurement terms are normal distributions (multivariate normal for $\mathbf{y}_i$) centered on the true values.
The function $h$ transforms the true heliocentric quantities into Galactocentric coordinates, so that the last term corresponds to the DF (Equation~\ref{eq:df}), which acts as the prior on the true values.
The true values are treated as latent parameters and are marginalized over by the sampler.
Stars lacking proper motion measurements are treated by marginalizing over this missing information with a broad prior on the proper motion components (see e.g., \citealt{Shen2022}).

We adopt the same hyperpriors on the model parameters as \citetalias{Medina2025c}:    
    
\begin{align}
  \alpha   &\sim \mathcal{N}(4.0,\; 0.1^2),   \nonumber \\
  \beta    &\sim \mathcal{N}(0.3,\; 0.05^2),  \nonumber \\
  (\Phi_0, \gamma) &\sim \mathcal{N}\!\left(
  \boldsymbol{\mu},\; \boldsymbol{\Sigma}
  \right),
  \label{eq:priors}
  \end{align}

  \noindent where $\boldsymbol{\mu} = (64.63,\; 0.440)$ and $\boldsymbol{\Sigma} = \bigl(\begin{smallmatrix} 47.61 & 0.165 \\ 0.165 & 0.0021 \end{smallmatrix}\bigr)$, adapted
  from \citet{Shen2022}.

Posterior sampling is performed via Hamiltonian Monte Carlo \citep[HMC;][]{Duane1987} with the No-U-Turn Sampler \citep[NUTS;][]{Hoffman2014}, as implemented in \textsc{Stan} \citep{Carpenter2017}. 
We run four independent chains, each with 1000 warm-up draws followed by 1000 post-warmup draws, and assess convergence using the $\hat{R}$ statistic \citep{Gelman1992}, requiring $\hat{R} \lesssim 1.05$ for all parameters.

\section{Method validation: Application to simulated galaxies}
\label{sec:mocks}

We validate our model by applying it to mock stellar catalogs constructed from cosmological simulations of Milky Way-mass galaxies.
This step is essential to characterize potential biases and systematic uncertainties in the mass recovery, given the strong assumptions of our model. 
In fact, spherical symmetry and dynamical equilibrium are not guaranteed to hold in realistic halos (particularly for tracers spanning broad distance ranges), and the assumptions of a power-law behavior of $\phi$ and $\rho$, in addition to the assumption of a constant velocity anisotropy in the halo, 
are strong oversimplifications. 
Indeed, tests on cosmological simulations have shown that DF fits assuming a fixed functional form, such as the $L^{-2\beta}f(\mathcal{E})$ form adopted by GME, can yield significantly biased mass estimates \citep{Han2016}, and that steady-state models applied to halo stars are subject to larger stochastic biases than when applied to satellite galaxies (as halo stars are less phase-mixed; see \citealt{Han2020}).

We draw mock samples from AuriDESI \citep{Kizhuprakkat2024}, a set of synthetic DESI MWS catalogs constructed from the Auriga magnetohydrodynamic zoom-in simulations \citep{Grand2017}.
AuriDESI self-consistently models stellar and gas physics and therefore offer realistic stellar halo populations (in a cosmological context) with full survey observational realism (target selection, spectral pipeline outputs, and measurement uncertainties). 
Furthermore, AuriDESI provides samples of observed BHB and RRL stars accounting for the selection function of DESI DR2 (based on both on-sky position and magnitude-dependent completeness).
A thorough description of the AuriDESI BHB and RRL catalogs used in this work 
is provided in a companion paper \citep[][]{Riley:2026auridesi}.

With respect to \citetalias{Medina2025c}, who validated the GME method on two AuriDESI mocks (Au-6 and Au-21) at a single viewing angle of $30\degr$, here we significantly expand the validation by considering six Auriga halos (Au-6, Au-16, Au-21, Au-23, Au-24, and Au-27) at four viewing angles each ($30\degr$, $120\degr$, $210\degr$, and $300\degr$).
This broader set allows us to assess the robustness of the method across a more representative range of halo morphologies, substructure levels, and line-of-sight orientations.

\subsection{The AuriDESI mock BHB and RRL catalogs}
\label{sec:auridesi}

We use mock catalogs from six Auriga halos: Au-6, Au-16, Au-21, Au-23, Au-24, and Au-27. 
As explained in more detail by \citet{Riley:2026auridesi}, these halos span a representative range of assembly histories and halo morphologies. 
Au-6 is the closest analog to the Milky Way in terms of stellar mass, star formation rate, and disk morphology, with a virial mass of $1.01 \times 10^{12}\,M_{\odot}$ and a quiescent recent accretion history \citep{Grand2017,Monachesi2019, Kizhuprakkat2024,Pu2025,Riley2025}. 
At the other extreme, Au-21, Au-23, and Au-27 have prominent ongoing interactions with massive satellites.
Au-21, for instance, hosts a massive satellite near $\sim\!60$\,kpc that produces strong kinematic substructure in the outer halo. 
In this regard, Au-21 is a closer analog of the ongoing interaction between the Milky Way and the LMC.
Au-16 and Au-24 are intermediate cases with more extended disks but no dominant active merger. 
These six halos span virial masses in the range  $1.0$--$1.7 \times 10^{12}\,M_{\odot}$, broadly consistent with current estimates of the Milky Way's total mass \citepalias[see Figure 11 in][]{Medina2025c}. 
 
For each halo, AuriDESI provides four mock catalogs corresponding to different angular positions of the solar observer in the galactic plane ($30\degr$, $120\degr$, $210\degr$, and $300\degr$ relative to the bar major axis), with the Sun placed at a Galactocentric radius of $8$\,kpc. 
This yields a total of 24 mock realizations which, compared to the two used by \citetalias{Medina2025c} (Au-6 and Au-21, at $30\degr$ only), allows us to further assess systematic effects arising from the specific line-of-sight orientation of the observer.

\subsubsection{Sample selection, distance, and velocity of \\BHBs and RRLs in AuriDESI}
\label{sec:mockspatial}

The AuriDESI mocks catalogs provide separate BHB and RRL samples for each of the 24 halo--viewpoint combinations described in Section~\ref{sec:auridesi}. 
BHBs and RRLs in the mock catalogs are identified from simulated horizontal branch populations using a temperature boundary at $T_{\rm eff} = 7{,}000$\,K, with hotter stars classified as BHBs and cooler stars as RRLs. 
This definition provides more details than the broader horizontal branch selection of \citet{Kizhuprakkat2024}, which included a contaminating population of young stars preferentially located in massive satellites. 
The resulting mock catalogs contain $\sim\!9{,}000$--$15{,}000$ BHBs and $\sim\!10{,}000$--$18{,}000$ RRLs across the 24 halo--viewpoint combinations before selection cuts, 
which are later divided into random subsets (Section~\ref{sec:mockcuts}).

Heliocentric distances in the mock catalogs are assigned a uniform uncertainty of 5\% for both BHBs and RRLs, consistent with the typical observational uncertainties in the DR2 samples (Section~\ref{sec:distances}). 
Line-of-sight velocity uncertainties are treated differently for each tracer. 
For BHBs, uncertainties are drawn from a magnitude--color grid calibrated to the DESI DR2 performance, which results in AuriDESI BHBs having slightly larger typical uncertainties ($\sim\!1.5$--$4.5\,{\rm km\,s^{-1}}$) than their observed counterparts.
This reflects the fact that mock BHBs tend to lie at larger Galactocentric distances and therefore fainter apparent magnitudes. 
For RRLs, uncertainties are sampled directly from the empirical distribution of DESI DR2 RRL velocity uncertainties \citep{Medina2025a}, which has a broad tail extending beyond $10\,{\rm km\,s^{-1}}$. 
Proper motion uncertainties are assigned based on {\it Gaia} DR3 performance models as a function of apparent $G$-band magnitude. 
Full details of the error model are given in \citet{Riley:2026auridesi}.

Two notable differences are observed between the DESI and the AuriDESI BHB and RRL samples. 
Firstly, the RRL-to-BHB ratio in the mocks is approximately unity out to 70\,kpc, while in the real data it is approximately unity out to 50\,kpc and then drops near 50\,kpc.
Secondly, the AuriDESI tracers are less centrally concentrated than the real DESI samples: roughly 50\% of mock tracers fall within 30\,kpc compared to $\sim\!80$\% for the real data.
Thus, these offsets should be taken into account when interpreting the mock validation results.

\subsubsection{Selection cuts on the mock BHB and RRL samples}
\label{sec:mockcuts}

Similar to \citetalias{Medina2025c}, we remove stars belonging to bound substructures by excluding particles with \texttt{pop\_id}$\,<\!2$ prior to applying our model. 
In addition, we remove stars associated with ongoing massive merger events ($>10^8$\,M$_\odot$), which are observed as massive stellar streams in every mock halo except for Au-6. 
We then apply to each mock catalog the same selection cuts used for the DESI DR2 data (Section~\ref{sec:cuts}), to ensure validation tests that replicate the conditions of our analysis on DESI observations. 

After applying the selection cuts of Section~\ref{sec:cuts} and removing bound substructure members, 
the AuriDESI BHB and RRL mocks contain $\sim\!1{,}600$--$2{,}700$ and $\sim\!2{,}000$--$3{,}800$ stars per realization, respectively.
As in \citetalias{Medina2025c}, the large number of tracers per mock makes the direct application of the GME computationally prohibitive, and we therefore adopt a divide-and-conquer approach: each mock catalog is randomly partitioned into 24 non-overlapping subsets.
Then, the GME is run independently on each subset, and the final mass estimate for a given halo--viewpoint combination is 
taken as the median of the pooled subset posteriors, with uncertainties given by their 16th and 84th percentiles. 
Each subset thus contains $\sim\!70$--$160$ stars, comparable to the subsets of the DESI DR2 samples (Section~\ref{sec:desiY3-models}), so that the mock calibration is performed in a similar regime as the analysis of the real data.

\subsection{Recovery of the mass of simulated galaxies}
\label{sec:mockmass}

We apply the GME method to each of the 24 AuriDESI mock realizations (six halos, each with four viewing angles) for both the BHB and RRL tracer samples (separately), and compare the recovered enclosed mass profiles with the true mass profiles measured directly from the simulations.

\begin{figure*}
        \centering    
        \includegraphics[width=0.49\textwidth]{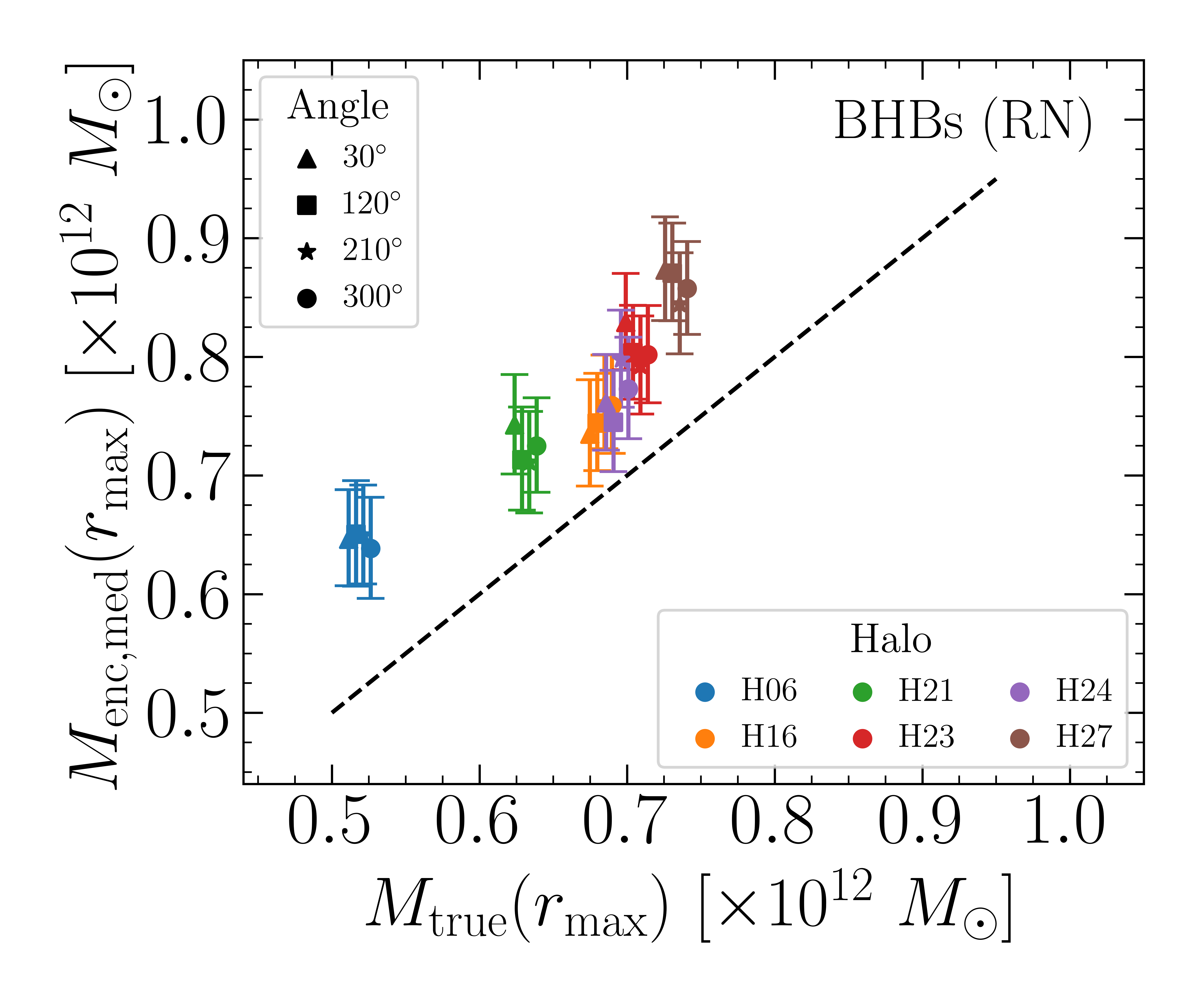}
        \includegraphics[width=0.49\textwidth]{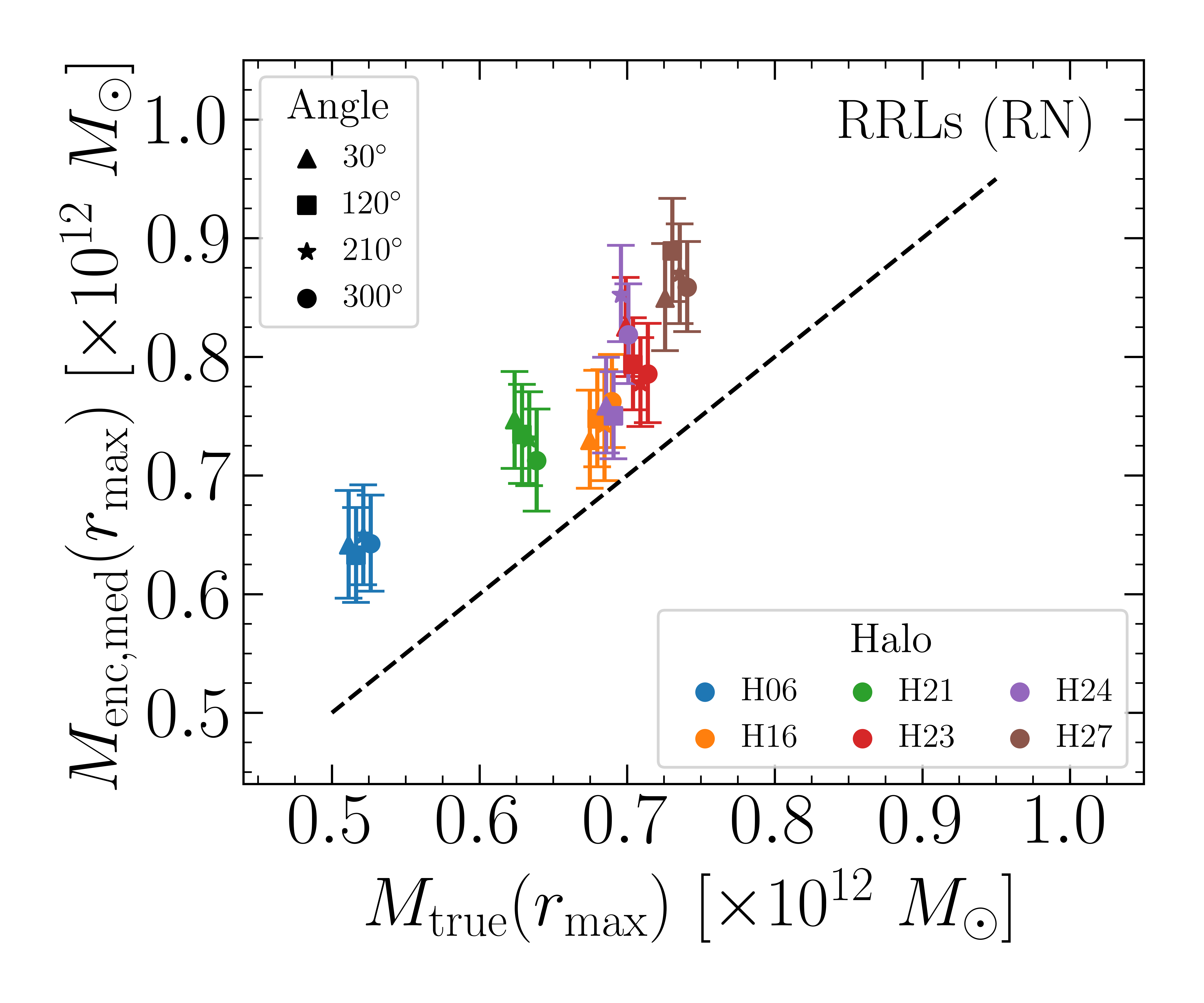}
         \includegraphics[width=0.49\textwidth]{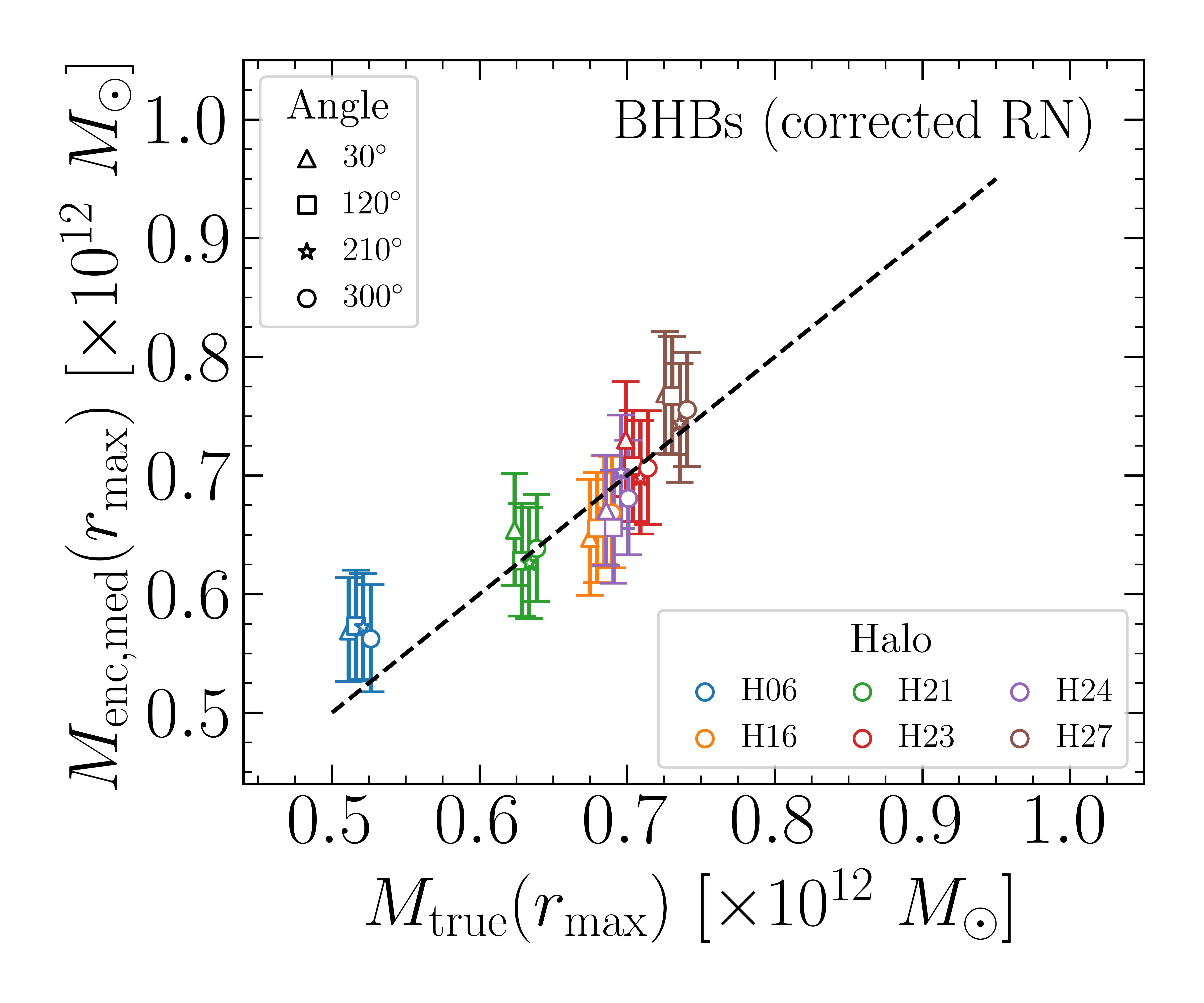}
        \includegraphics[width=0.49\textwidth]{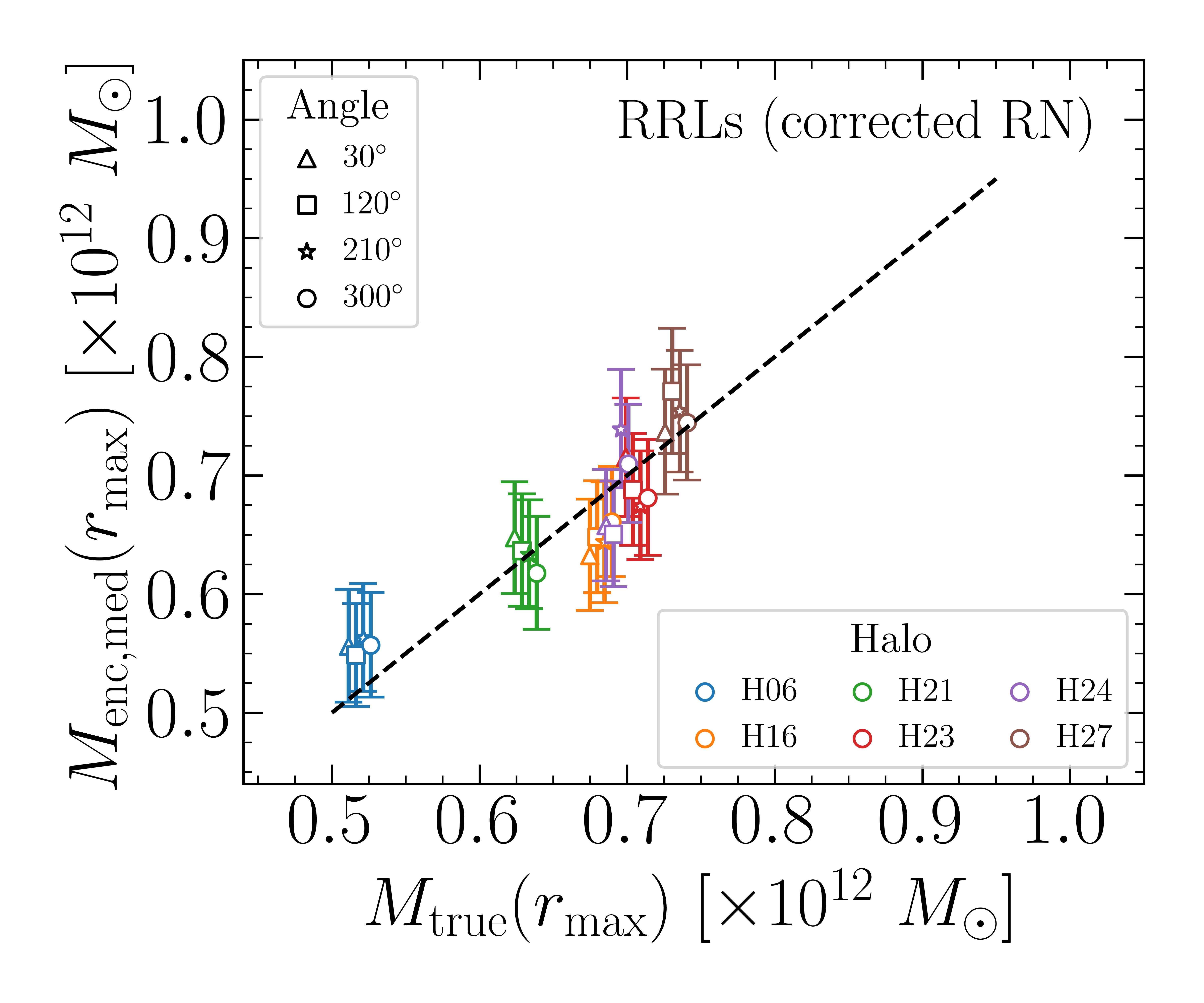}
    \caption{Recovered versus true enclosed mass at the maximum Galactocentric radius $r_{\rm max}$ of each mock sub-sample, for the 24 AuriDESI halo--viewpoint combinations (6 halos $\times$ 4 viewing angles). \textit{Top:} GME model with radial normalization (Section~\ref{sec:model}). 
\textit{Bottom:} GME model with radial normalization and bias correction applied.  
\textit{Left and right} panels show results for BHB and RRL tracers, respectively. 
Each point represents the median recovered mass for a given halo--viewpoint combination, with error bars representing the 16th--84th percentiles of the 
pooled
posterior.
Colors identify the six Auriga halos and symbols identify the four viewing angles ($30^{\circ}$, $120^{\circ}$, $210^{\circ}$, $300^{\circ}$). 
The dashed line indicate the identity function. 
Our model systematically overestimates the enclosed mass across all halos and viewing angles
The computation of this systematic bias  
is discussed further in Section~\ref{sec:mockmass}.
    }
    \label{fig:mass_recovery}
\end{figure*}

\begin{figure*}
        \centering    
        \includegraphics[width=0.49\textwidth]{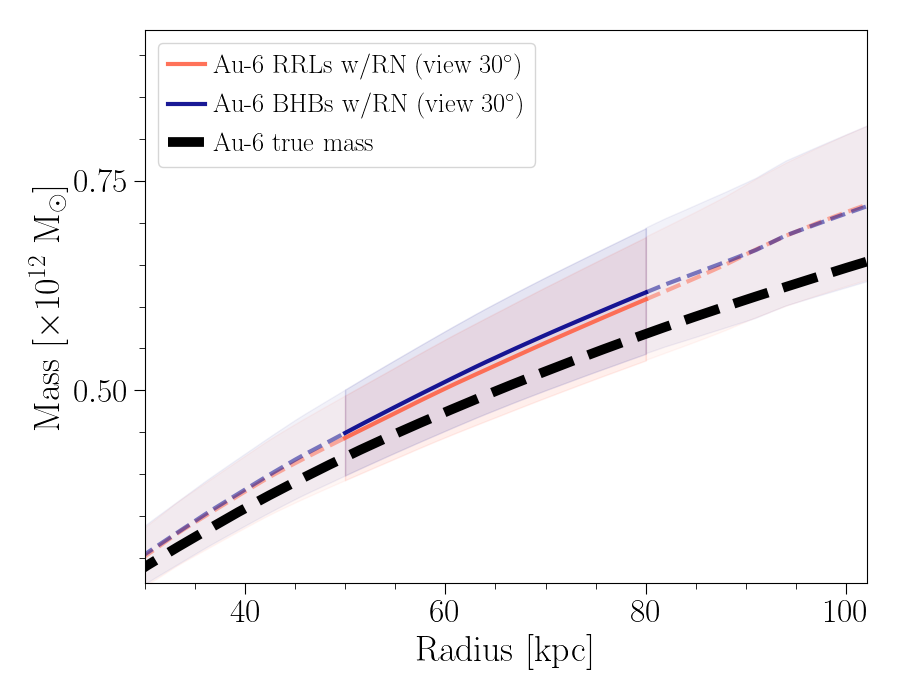}
        \includegraphics[width=0.49\textwidth]{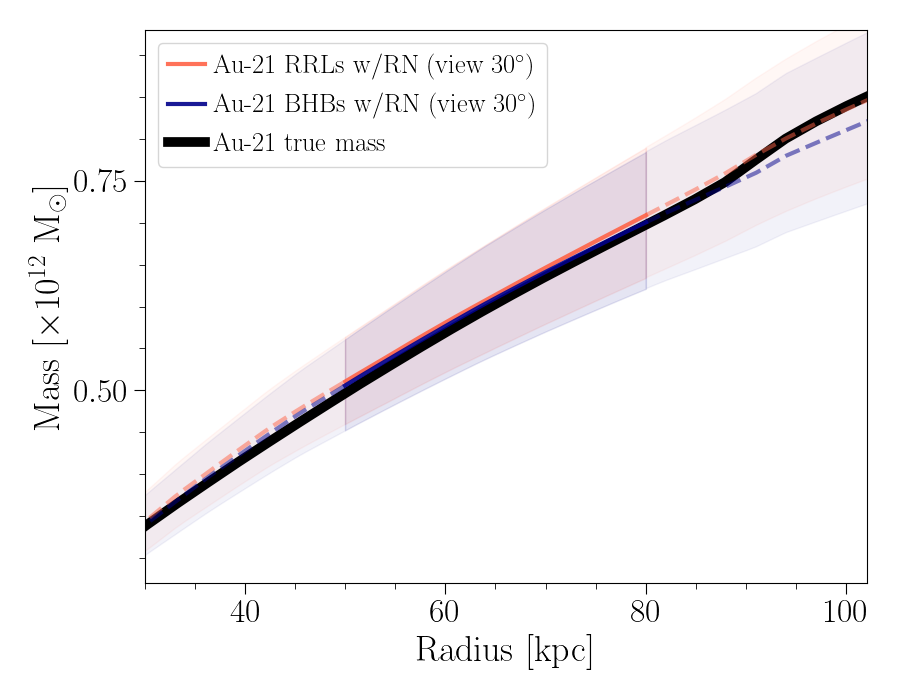}
        \includegraphics[width=0.49\textwidth]{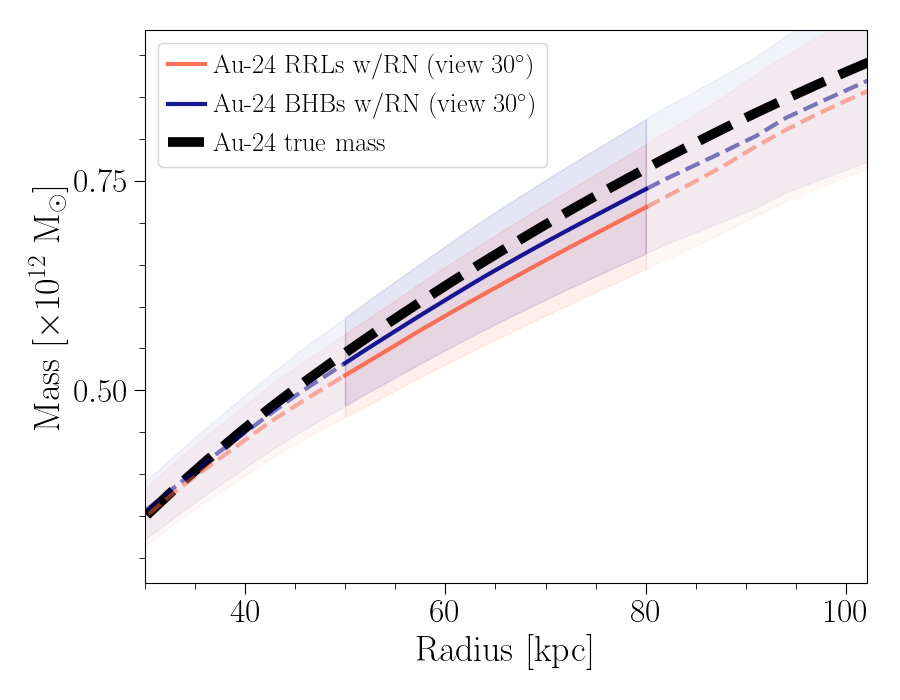}
    \caption{
    Cumulative mass profiles (CMPs) recovered from mock BHB (blue) and RRL (orange) catalogs for three AuriDESI mock halos (Au-6, Au-21, and Au-24) at a fixed viewing angle of $30^{\circ}$, compared with the true enclosed mass of the simulation (black dashed lines). 
    These results were obtained with radial normalization (RN) applied, and corrected by the overall AuriDESI systematic bias (see Section~\ref{sec:mockmass}).
    Shaded regions represent the confidence intervals computed from the 2.5, 12.5, 25.0, 75.0, 87.5, and 97.5 percentiles of the enclosed mass distribution. Solid curves indicate the range covered by the mock tracers; dashed curves show the extrapolation beyond that range. Vertical lines mark the minimum and maximum Galactocentric distances of the mock tracer samples. 
    Before correction, both tracers overestimate the true enclosed mass by $\sim\!15\%$ at $r = 80\,\mathrm{kpc}$ for all halos (see Figure~\ref{fig:mass_recovery}). 
    After applying the correction, the recovered CMPs match the true mass profiles within uncertainties. 
    }
    \label{fig:mass_recovery-halos6-21-24}
\end{figure*}

A quantitative view of the mass recovery is shown in Figure~\ref{fig:mass_recovery}. The top panels show results obtained with radial normalization applied.
The recovered masses are systematically overestimated across all halos and viewing angles, by $\sim\!15$\% on average at $r_{\rm max}=80$\,kpc for both BHBs and RRLs.

To translate the mock-calibrated overestimation into a correction applicable to real data, we compute, for each of the 24 AuriDESI realizations, the ratio $f(r) = M_{\rm true}(<r)/M_{\rm enc}^{\rm RN}(<r)$ as a function of Galactocentric radius, where $M_{\rm true}$ is the true simulation enclosed mass and $M_{\rm enc}^{\rm RN}$ is the GME-recovered median with RN applied. We then derive the median correction factor $\langle f(r)\rangle$ and its 1$\sigma$ scatter $\sigma_f(r)$ across the 24 realizations, and use these in Section~\ref{sec:desiY3-models} to produce a corrected DESI mass profile. 
The individual $f(r)$ curves of the 24 realizations, together with $\langle f(r)\rangle$ and $\sigma_f(r)$, are shown in Figure~\ref{fig:fr_all24}. 
In all realizations $f(r)<1$ across the radial range of the data, with $\langle f(r)\rangle\sim\!0.85$ and $\sigma_f\simeq0.04$ between 50 and 80\,kpc.
For an illustrative purpose, we show the effect of correcting our mass estimates in AuriDESI by $f(r)$ in Figure~\ref{fig:mass_recovery}.

As a representative example, 
Figure~\ref{fig:mass_recovery-halos6-21-24} 
displays the 
corrected 
cumulative mass profile (CMP), $M(<r)$, for three AuriDESI halos (Au-6, Au-21, and Au-24)  at a viewing angle of $30\degr$.  
Before correction, both BHB and RRL tracers overestimate the true enclosed mass by $\sim\!15$\% at $r = 80$\,kpc (Figure~\ref{fig:mass_recovery}).
As shown in Figure~\ref{fig:mass_recovery-halos6-21-24}, applying the systematic bias correction $f(r)$ makes the CMPs match the true mass profiles of the halos (within uncertainties).

Since all 24 realizations are biased in the same direction and display a similar radial dependence (Appendix~\ref{sec:fr_all}), a single median correction applies to the full mock set, rather than only to the three halos shown here.
When extending the analysis across all 24 mock realizations, we find that 
larger 
deviations occur in halos with prominent ongoing mergers (Au-21, Au-23, Au-24, Au-27), where the assumption of dynamical equilibrium does not strictly hold.

We examine the effect of retaining stars associated with massive stellar streams on the recovered masses, noting in passing that Au-6 lacks massive satellite interactions (hence this effect is not tested on that halo). 
Retaining stream members has no significant systematic effect on the recovered masses in the mocks ($\lesssim\!1$\% on average at $r = 80$\,kpc across all halos and viewing angles).
However, a residual overestimation persists in the most active halos even after stream removal, reflecting the fact that their kinematics are not fully captured by an equilibrium model. This residual bias is discussed further in Section~\ref{sec:selection_cuts_biases}.

Additionally, we assess the effect of velocity cuts on the mock mass recovery for all six halos at a single viewing angle, in combination with the RN model. 
We find that adopting azimuthal ($v_T$) and tangential ($v_{\rm tan}$) cuts at 500\,km\,s$^{-1}$ (separately) yield broadly consistent results and 
partially reduce this overestimation
(by $\sim\!5$\%), but do not eliminate it. 
The impact of these cuts on the real DESI DR2 estimates is discussed in Section~\ref{sec:selection_cuts_biases}.

\section{Galactic mass estimation with DESI DR2}
\label{sec:desiY3-models}

We apply the GME method (Section~\ref{sec:model}), tested on AuriDESI mock samples, to the BHB and RRL samples described in Section~\ref{sec:data}. 
Given the substantially larger sample sizes relative to \citetalias{Medina2025c} (and the computational cost of sampling the per-star latent parameters), we adopt a divide-and-conquer strategy, 
dividing each sample into two independent subsets. 
Each subset is sampled using 
NUTS, 
adopting the same setup as \citetalias{Medina2025c} and as described in Section~\ref{sec:hierarchicalbayesianmodel}.
The posterior is sampled separately for each subset, and the final enclosed mass profile is derived from the pooled posterior draws of the two subsets, as done for the mocks (Section~\ref{sec:mockcuts}).
Since pooling does not fully combine the constraining power of the two subsets, the uncertainties we report are conservative. 
As a reference, sampling on a single workstation a set of $\sim100$ tracers with four parallel chains of 1000 warm-up and 1000 post-warm-up draws each takes $\sim$20 minutes (where the run time is dominated by the per-star latent parameters).
We find that $\hat{R} \lesssim$ 1.05 are typically reached by this configuration.
Our reported values correspond to the median of the distributions and the $1\sigma$ uncertainties to the 16th--84th percentile intervals.

The results presented in this section are obtained with the RN modification described in Section~\ref{sec:model}. 
As demonstrated by our mock validation in Section~\ref{sec:mockmass}, 
our model systematically overestimates the enclosed mass by $13$--$20\%$ 
across all halos and viewing angles. 
To account for this bias, we apply the mock-calibrated correction factor $\langle f(r)\rangle$ derived in Section~\ref{sec:mockmass}, 
which makes the corrected enclosed mass to be
$M_{\rm corr}(<r) = M^{\rm RN}(<r)\times \langle f(r)\rangle$, 
with a propagated uncertainty that accounts for the 
1$\sigma$ scatter of the correction factor across the 24 mock realizations. 
We report $M_{\rm corr}$ as our primary result.

Figure~\ref{fig:cornerplots} shows the posterior distribution of our model parameters, computed from our RRL and BHB samples.
The potential slope inferred by our model is $\gamma = 0.46^{+0.05}_{-0.06}$ for RRLs and $\gamma = 0.45^{+0.04}_{-0.05}$ for BHBs, consistent with a mildly declining circular velocity curve in the outer halo. 
The potential normalization is $\Phi_0 = 53.1^{+7.3}_{-7.2} \times 10^4\,\mathrm{km^2\,s^{-2}\,kpc^{\gamma}}$ (RRLs) and $\Phi_0 = 62.1^{+6.7}_{-6.4} \times 10^4\,\mathrm{km^2\,s^{-2}\,kpc^{\gamma}}$ (BHBs).
The orbital anisotropy is $\beta = 0.33^{+0.05}_{-0.05}$ for RRLs and $\beta = 0.34^{+0.05}_{-0.05}$ for BHBs.
These results represent halo velocity distributions that are slightly more radial than those obtained by \citetalias{Medina2025c} using DESI DR1 
($0.30^{+0.05}_{-0.05}$ and $0.25^{+0.04}_{-0.04}$  for RRLs and BHBs, respectively), but are in a better agreement with previous $\beta$ estimates in the literature at similar distances, e.g.,  \citet{Bird2019} and \citet{Shen2022} using giant stars ($<0.40$ and $0.36^{+0.04}_{-0.04}$, respectively), and \citet{Medina2025a} using RRLs in DESI DR1 and accounting for the presence of stars from the Gaia-Sausage-Enceladus merger event ($\sim\!0.35$).
The inferred density slope is $\alpha = 4.57^{+0.08}_{-0.08}$ for RRLs and $\alpha = 4.71^{+0.08}_{-0.08}$ for BHBs. 
Both values are consistent with the steep outer-halo profiles measured directly from the spatial distribution of these same tracer populations at comparable Galactocentric distances for RRLs \citep[e.g.][]{Sesar2011, Iorio2021, Medina2018,Medina2024} and for BHBs  \citep[e.g.][]{Deason2011b, Thomas2018}.

\begin{figure}
        \centering    
        \includegraphics[width=0.49\textwidth]{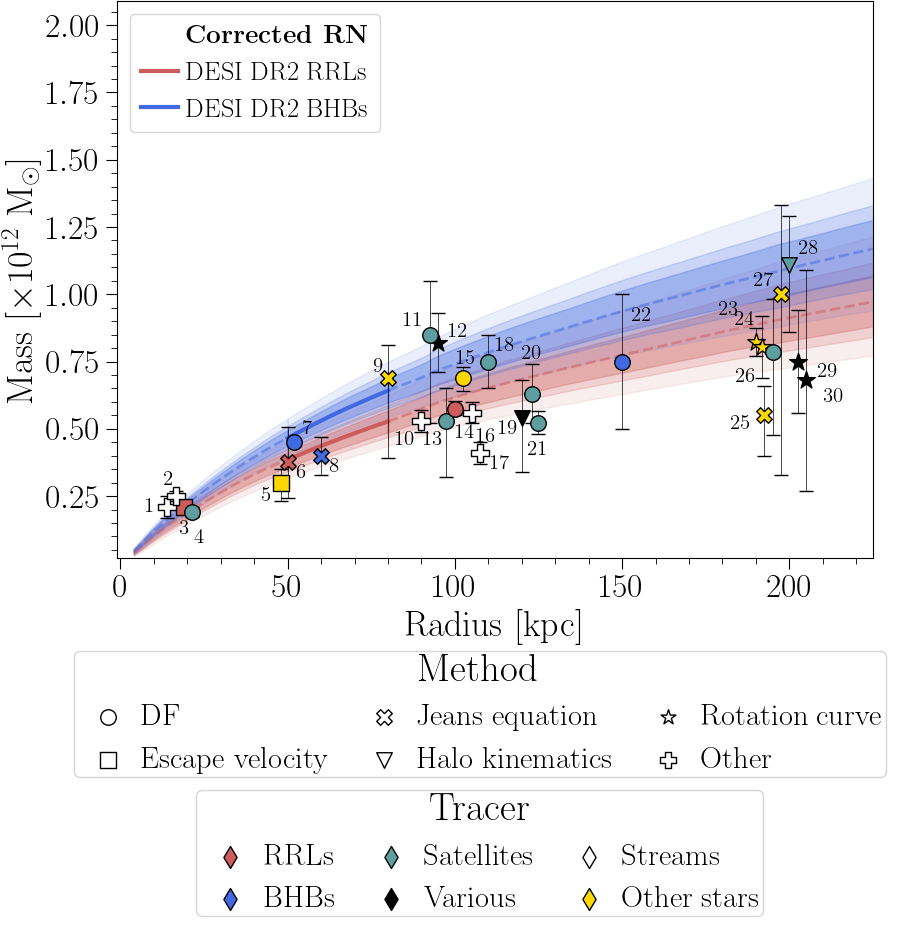}
    \caption{
    CMPs of the Galaxy derived from our BHB and RRL samples (solid curves), compared with enclosed mass estimates from the literature (numbered symbols). Shaded regions represent the confidence intervals computed from the 2.5, 12.5, 25.0, 75.0, 87.5, and 97.5 percentiles of the enclosed mass distribution. Vertical markers indicate the Galactocentric distances of the individual tracers used for our mass modelling. 
    The results shown are obtained with our radial normalization (RN) method and correcting for the systematic bias observed in the AuriDESI mocks. 
    Similar to Figure~\ref{fig:mass_recovery-halos6-21-24}, solid curves indicate the radial range covered by our tracers and dashed curves show the extrapolation of the fitted power-law profiles.
    Numbered literature estimates follow:
1--\citep{Kuepper2015},
2--\citep{Malhan2019},
3--\citep{Prudil2022},
4--\citep{Posti2019},
5--\citep{Williams2017},
6--\citep{Ablimit2017},
7--\citep{Williams2015},
8--\citep{Xue2008},
9--\citep{Gnedin2010},
10--\citep{Bayer2025},
11--\citep{Vasiliev2019},
12--\citep{McMillan2017},
13--\citep{Eadie2019},
14--\citep{Hattori2021},
15--\citep{Shen2022},
16--\citep{Vasiliev2021},
17--\citep{Gibbons2014},
18--\citep{CorreaMagnus2022},
19--\citep{Battaglia2005},
20--\citep{Eadie2017},
21--\citep{Eadie2016},
22--\citep{Deason2012b},
23--\citep{Ablimit2020},
24--\citep{Zhou2023},
25--\citep{Bird2022},
26--\citep{Wang2022},
27--\citep{Bird2022},
28--\citep{Sun2023},
29--\citep{Bajkova2016},
30--\citep{Bhattacharjee2014}.
Small shifts in distance were added to the markers at 20, 50, 100, 125, and 200\,kpc to improve the readability of the comparison. These shifts range between 1.5 and 5\,kpc, depending on the level of crowding at a given distance. 
    }
    \label{fig:CMP-rrlbhb}
\end{figure}

Figure~\ref{fig:CMP-rrlbhb} shows the inferred enclosed mass profile $M(<r)$ as a function of Galactocentric radius for both tracers, after correcting for the systematic bias observed in our AuriDESI mocks. 
The BHB profile lies systematically above the RRL profile across the radial range probed by our samples, although the two agree within $\sim\!1.5\sigma$ overall.
At 80\,kpc, close to the outermost radius of each sample, we obtain the following enclosed masses:

\begin{align}
M(<80\,{\rm kpc}) &=
0.64^{+0.06}_{-0.06} \times 10^{12}\,M_\odot
\quad \text{(BHBs)}, \\
M(<80\,{\rm kpc}) &=
0.53^{+0.05}_{-0.05} \times 10^{12}\,M_\odot
\quad \text{(RRLs)}.
\end{align} 

For a direct comparison with the DR1 estimates of \citetalias{Medina2025c} and with other results in the literature, we also report the enclosed mass at $100$\,kpc:

\begin{align}
M(<100\,{\rm kpc}) &= 0.74^{+0.07}_{-0.07} \times 10^{12}\,M_\odot \quad \text{(BHBs)}, \\
M(<100\,{\rm kpc}) &= 0.62^{+0.06}_{-0.07} \times 10^{12}\,M_\odot \quad \text{(RRLs)}.
\end{align}

We caution that $M(<100 {\rm kpc})$ is obtained by extrapolating 
our cumulative mass profile beyond the extent of our data, under the assumption that our model is also valid in that distance range.
Therefore, the reliability of this extrapolation is not guaranteed. 
A detailed comparison of our $M(<100\,{\rm kpc})$ estimates with our previous results (using DESI DR1) is given in Section~\ref{sec:dr1comparison}.

In addition, we estimate the virial mass of the Milky Way, $M_{200}$, by extrapolating our GME constraints beyond the radial range covered by our data. 
We note that extrapolating the fitted power-law potential directly out to the virial radius ($R_{200}\gtrsim220$\,kpc) is, in general, not a valid approach, as it implicitly assumes that our model's underlying assumptions hold at radii far beyond those probed by our tracers.
We instead treat our GME result as a local measurement of the enclosed mass and its logarithmic slope at the anchor radius $r_0=65$\,kpc (the midpoint of our fitted range, 50–80\,kpc), and fit a physically motivated NFW halo with a fixed baryonic  disk and bulge component to each posterior sample of $(\Phi_0,\gamma)$. 
Our results, obtained from a propagation of the full posterior chain through this mapping yields the $M_{200}$ values provided below:

\begin{align}
M_{200} &= 1.12^{+0.14}_{-0.12} \times 10^{12}\,M_\odot \quad \text{(BHBs)}, \\
M_{200} &= 0.89^{+0.11}_{-0.10} \times 10^{12}\,M_\odot \quad \text{(RRLs)}.
\end{align}

The complete formalism followed for this derivation is described in Appendix~\ref{sec:m200_computation}.

\section{Discussion}
\label{sec:discussion}

We have applied the GME method to 279 BHBs and 213 RRLs from DESI DR2, obtaining enclosed mass profiles extending to $\lesssim80$\,kpc from the Galactic center. 
In the following subsections we assess the robustness of these estimates against systematic effects arising from our model and selection choices (Section~\ref{sec:selection_cuts_biases}), compare them with the DESI DR1 results of \citetalias{Medina2025c} (Section~\ref{sec:dr1comparison}), 
and place them in the context of the broader literature (Section~\ref{sec:comparisonLiterature}).

\subsection{Effect of the selection cuts and other biases}
\label{sec:selection_cuts_biases}

We assess the sensitivity of our mass estimates to our modeling and selection choices, 
namely the mock-based bias correction (Section~\ref{sec:mockmass}) and the removal of Sgr stream members.
Additionally, we discuss the effect of high-velocity stars (potentially unbound tracers or contaminants) on our mass estimates.

The dominant systematic effect in our analysis is the overestimation of the enclosed mass by our model 
across all AuriDESI halos and viewing angles (by 13--20\%; Section~\ref{sec:mockmass} and Figure~\ref{fig:mass_recovery}). 
We account for this bias through the mock-calibrated correction factor 
$f(r)$ 
(Section~\ref{sec:desiY3-models}), which absorbs, to first order, the effects that are not included in the normalization (e.g., the flux-limit truncation of our samples; see Section~\ref{sec:model}).

Including Sgr stream members 
lowers 
the inferred enclosed mass by 
9--11\% for both BHBs and RRLs 
at $r = 80$\,kpc, relative to the Sgr-removed baseline. 
We note that this effect is not reproduced in the AuriDESI mocks, where retaining stream stars has no significant systematic effect on 
the recovered masses.
Thus, this discrepancy is likely explained by the completeness and purity of the selection of Sgr members in our real data (which relies on the  positions and distances of the tracer samples), or by the fraction of Sgr stars $>50$\,kpc being larger than the fraction of stars in similarly massive streams in AuriDESI. 
Conversely, removing all stars within the region where Sgr members are selected (i.e., within $15^\circ$ from the Sgr track in stream latitude, as described by \citetalias{Medina2025c}; Section~\ref{sec:cuts}), regardless of distance, avoids the distance-dependent gaps left by the removal based on our fiducial membership selection. 
This reduces the samples by 44\% (RRLs) and 33\% (BHBs), but increases the enclosed masses within 50--80\,kpc by only 5\% and 3\%, respectively ($\lesssim0.5\sigma$).

Applying a $500\,{\rm km\,s^{-1}}$ cut on either $v_T$ or $v_{\rm tan}$, similar to \citetalias{Medina2025c}, reduces the inferred enclosed mass relative to the no-velocity cut baseline (when tested without using RVS\_WARN to remove sources from our samples), though with a notable tracer dependence. 
For RRLs, neither velocity cut has a significant effect at $r = 80$\,kpc ($\lesssim\!1$\% for both). 
For BHBs, the tangential cut is more aggressive, lowering the mass by $\sim\!20\%$, whereas the azimuthal cut lowers it by $\sim\!13\%$. 
This asymmetry is likely explained by the different kinematic distributions of BHBs and RRLs in our sample (as seen in Figure~\ref{fig:dist_and_velocity_distribution}).
Our BHB sample contains a population of stars with high total $v_{\rm tan}$ but only moderate azimuthal velocity, making the tangential cut particularly aggressive for that tracer. 
In contrast, the RRL sample displays no significant high-velocity tail beyond the cut threshold, so neither cut meaningfully affects the RRL mass estimate. 
For BHBs, the stars removed by the tangential cut are likely either contaminants with overestimated distances, and hence tangential velocities (Section~\ref{sec:cuts}), or stars on nearly radial orbits whose kinematics are not well captured by the equilibrium DF. 
Most of these stars are removed by our RVS\_WARN cut (Sections~\ref{sec:cuts} and ~\ref{sec:comparisonLiterature}).

\subsection{Comparison with DESI DR1 mass estimates}
\label{sec:dr1comparison}

In \citetalias{Medina2025c} we measured the enclosed Galactic mass using $330$ 
BHBs and $110$ RRLs in DESI DR1.
The DR2 RRL sample used in this work is roughly twice as large, while the BHB sample is of similar size. 
Although the parent DR2 BHB catalog is about twice as large as that of DR1, here we restrict the BHB sample to the \textsc{bright} program of the main DESI survey (\citetalias{Medina2025c} included BHBs from all DESI programs), applying also an additional cut on the quality flags of the spectroscopic analysis pipeline (Section~\ref{sec:cuts}).
As a consequence, the two BHB samples also differ in radial coverage, with the DR1 sample extending to $R_{\rm GC}\sim100$\,kpc and the DR2 sample limited to $R_{\rm GC}\lesssim80$\,kpc.
In this section, we compare our DR1 and DR2 results.

The current DR2 results are obtained with radial normalization, which was not employed in \citetalias{Medina2025c}.
At the reference radius of $100$\,kpc, \citetalias{Medina2025c} reported $M(<100\,{\rm kpc}) = 0.57^{+0.08}_{-0.07} \times 10^{12}\,M_\odot$ (BHBs) and $0.55^{+0.12}_{-0.10} \times 10^{12}\,M_\odot$ (RRLs), while in this work we obtain $M(<100\,{\rm kpc}) = 0.74^{+0.07}_{-0.07} \times 10^{12}\,M_\odot$ (BHBs) and $0.62^{+0.06}_{-0.07} \times 10^{12}\,M_\odot$ (RRLs). 
Thus, the BHB estimate is higher than the DR1 value by $\sim\!30$\%, while the RRL estimate is higher by $\sim\!13$\%. 
The fractional uncertainty on $M(<100\,{\rm kpc})$ is 
$\sim\!25$\% smaller than in DR1 for BHBs 
and $\sim\!50$\% smaller for RRLs, 
reflecting the significantly larger RRL sample size with respect to DESI DR1.

\subsection{Comparison with the literature} 
\label{sec:comparisonLiterature}

Figure~\ref{fig:CMP-rrlbhb} compares the enclosed mass profiles derived from our DR2 BHB and RRL samples with the compilation of enclosed mass estimates from the literature used by \citetalias{Medina2025c}, spanning Galactocentric radii from $\sim\!10$ to $>\!220$\,kpc. The literature values are drawn from a broad range of methods and tracers. 
These methods include distribution function modelling \citep[e.g.,][]{Eadie2019, Shen2022}, Jeans equations \citep[e.g.,][]{Gnedin2010}, escape velocity analyses \citep[e.g.,][]{Prudil2022}, and rotation curve fitting \citep[e.g.,][]{Zhou2023}, and tracers such as halo stars, globular clusters, dwarf satellite galaxies, and stellar streams. The figure shows results for our nominal configuration (i.e., with radial normalization and AuriDESI-based correction; see Section~\ref{sec:desiY3-models}).

At $r = 100$\,kpc, our corrected RN profiles yield  $M(<100\,{\rm kpc}) = 0.74^{+0.07}_{-0.07}$ and $0.62^{+0.06}_{-0.07} \times 10^{12}\,M_\odot$, for BHBs and RRLs respectively. 
Both estimates fall within the range spanned by the literature at this radius, with the BHB estimate lying above the RRL one (with slightly overlapping $1\sigma$ regions).
We note that the biases resulting from applying our selection cuts (Section~\ref{sec:selection_cuts_biases}) should be considered when interpreting the comparisons. 

Our CMPs are broadly consistent with several independent estimates in the literature. 
The RRL profile is in good agreement (within uncertainties) with the results of \citet{Shen2022}, who obtained $M(<100\,{\rm kpc}) = 0.69 \times 10^{12}\,M_\odot$ using halo giant stars from the H3 survey and \textit{Gaia} EDR3 data with a similar distribution function approach. Our RRL profile is also consistent  with \citet{Eadie2019}, who used globular clusters as tracers and found a mass within $125$\,kpc of $0.52 \times 10^{12}\,M_\odot$ and a virial mass $M_{200} = 0.68 \times 10^{12}\,M_\odot$. At larger radii, beyond $\sim\!80$\,kpc where our mass estimates rely on extrapolation of the power-law potential, comparison with studies based on tracers that directly probe the outer halo (in particular, globular clusters and dwarf satellite galaxies; see e.g., \citealt{Watkins2019, Posti2019}), become less informative. 
Our BHB profile also lies within the scatter of most outer-halo constraints across the distance range probed by our data.

We find that for BHBs using the quality cut on RVS\_WARN described in Section~\ref{sec:cuts} yields similar enclosed masses as 
applying a $v_{\rm tan}$ cut similar to that adopted by \citetalias{Medina2025c} ($v_{\rm tan} < 500\,{\rm km\,s^{-1}}$).
Indeed, such velocity cut applied to our sample with the same cuts but ignoring RVS\_WARN  
reduces the BHB enclosed mass at $100$\,kpc by $\sim\!20$\% ($0.69^{+0.06}_{-0.06} \times 10^{12}\,M_\odot$), bringing it into closer alignment with the center of the literature distribution. 
Conversely, the RRL estimate is essentially unchanged ($\lesssim\!1$\%, consistent with our AuriDESI mock tests) when adopting a tangential velocity cut. 
Thus, the comparison between our results and the literature is 
particularly sensitive to the mock-based bias correction and to the presence of high 
$v_{\rm tan}$ for BHBs (or contaminants in the sample, in part flagged by RVS\_WARN), but only to the former for RRLs.

\section{Conclusions}
\label{sec:conclusions}

We have applied the Galactic Mass Estimator (GME) to 
large, homogeneous spectroscopic samples of outer-halo BHB and RRL stars 
drawn from the first three years of DESI operations (DR2). 
After selection cuts, the analysis samples comprise 279 BHBs and 213 RRLs extending to $\sim\!80$\,kpc from the Galactic center. 
The RRL sample is roughly twice as large as that used in our previous work, while the BHB sample is of similar size, as we now restrict it to the main survey \textsc{bright} program.
Throughout this analysis, we employ a modified version of the standard GME, for which we implement a radial normalization that enables a better modeling of the tracer's spatial distribution. 
Given the scale of these samples we adopt a divide-and-conquer strategy, splitting 
each catalog into two independent subsets and pooling their posterior draws to derive the final enclosed mass profile, which yields conservative uncertainties.

Before applying the method to the real data, we validated it on a significantly expanded set of mock stellar catalogs with respect to our previous work: six AuriDESI halos (Au-6, Au-16, Au-21, Au-23, Au-24, Au-27) at four viewing angles each.
This validation shows that our model systematically overestimates the enclosed mass
across all halos and viewing angles, by $\sim\!15$\% at $r = 80$\,kpc for both tracers, with a similar radial dependence in all realizations.
Our reported results are corrected by this systematic offset. 

We assessed the sensitivity of our mass estimates to 
modeling and selection
choices. 
The dominant systematic effect is the mock-calibrated overestimation of the enclosed mass by our model, which we correct for. 
Similar to our previous work, we 
remove candidate members of the Sgr stream 
as a standard step, and find that  
including stars from 
this massive merger debris
lowers 
the enclosed mass by $\sim\!10$\% for both RRLs and BHBs at $r = 80$\,kpc.
When applied without our RVS\_WARN cut, velocity cuts (either on azimuthal $v_T$ or tangential $v_{\rm tan}$ velocities, at $500$\,km\,s$^{-1}$) affect the two tracers differently. 
For BHBs, the tangential cut lowers the mass by $\sim\!20$\%, while the azimuthal cut lowers it by $\sim\!13$\%. 
For RRLs, neither cut has a significant effect ($\lesssim\!1$\%).
This discrepancy is consistent with the observed differences in the high-velocity tail distributions of the two populations. 
For BHBs, our RVS\_WARN quality cut yields masses similar to those obtained with the tangential velocity cut.
We adopt the configuration with radial normalization, with Sgr removal, and without cuts on $v_T$ or  $v_{\rm tan}$ as our fiducial analysis.

Our primary results are the mock-corrected enclosed mass profiles $M_{\rm corr}(<r) = M^{\rm RN}(<r)\times \langle f(r)\rangle$, where the correction factor $\langle f(r)\rangle$ is calibrated from the 24 AuriDESI realizations and 
directly accounts for the systematic overestimation found in the mocks.
Extrapolating our profiles to $100$\,kpc, we obtain $M(<\!100\ {\rm kpc}) = 0.74^{+0.07}_{-0.07} \times 10^{12}\,M_\odot$ (BHBs) and $0.62^{+0.06}_{-0.07} \times 10^{12}\,M_\odot$ (RRLs),
and virial masses of 
$M_{200} = 1.12^{+0.14}_{-0.12} \times 10^{12}\,M_\odot$ (BHBs) and
$0.89^{+0.11}_{-0.10} \times 10^{12}\,M_\odot$ (RRLs). 
Since the mock-based correction removes this bias, 
these values are comparable to the DR1 estimates of \citetalias{Medina2025c}, which did not employ radial normalization. The BHB estimate is $\sim\!30$\% higher than the DR1 value of $0.57\times10^{12}\,M_\odot$, while the RRL estimate is consistent within uncertainties with the DR1 value of $0.55\times10^{12}\,M_\odot$. 
Unlike \citetalias{Medina2025c}, where both tracers yielded consistent masses, we find that BHBs yield systematically higher enclosed masses than RRLs (by $\sim\!20$\% at 80\,kpc), 
and the statistical uncertainties for RRLs are $\sim\!50\%$ smaller than in DR1, reflecting the larger sample.

This paper is one of three independent Milky Way mass measurements carried out as part of the DESI Year 3 Key Project (Y3KP), all applied to the same BHB and RRL tracer populations using complementary methods: Jeans modeling \citep[][]{Valluri:2026nimble} and an empirical DF approach (\citealt{Zhou:2026empdf}; based on the emPDF method of \citealt{Li2025}).
Combined results are discussed in \citet{DESI:2026mwmass}. 
Additional improvements to the GME method, including a treatment of the survey selection function (e.g., the flux-limit truncation of the tracer samples, currently accounted for through our mock-based correction), 
a systematic identification of potentially unbound tracers, and a correction for the gravitational influence of the Large Magellanic Cloud, will be implemented in a forthcoming companion paper (Slizewski et al., in prep.).

\section*{Data availability}
The data shown in all the figures of this manuscript will be made available in Zenodo (\href{https://doi.org/10.5281/zenodo.23710246}{https://doi.org/10.5281/zenodo.23710246}).

\acknowledgments

We are grateful to Ewoud Wempe (Observatoire de Paris) for insightful discussions and helpful feedback on the models used in this work.
GEM acknowledges support from the Dutch Research Council (NWO) through the Veni grant VI.Veni.252.355. 
GEM and TSL acknowledge financial support from Natural Sciences and Engineering Research Council of Canada (NSERC) through grant RGPIN-2022-04794.
TMC and SK acknowledge support from STFC grant ST/Y001001/1.  
AHR was supported by a fellowship funded by the Wenner-Gren Foundations (Wenner-Gren Stiftelserna), a Research Fellowship from the Royal Commission for the Exhibition of 1851, and by the Science \& Technology Facilities Council (STFC) through grant ST/T000244/1.
MV acknowledges support from  NASA-ATP grants 80NSSC20K0509 and 80NSSC24K0938.   
JH is supported by NSFC 12595312 and National Key R \& D Program of China (2023YFA1607800, 2023YFA1607801). 
LBeS acknowledges support from CNPq (Brazil) through a research productivity fellowship, grant no. [304873/2025-0]
GME acknowledges financial support from NSERC through grant RGPIN-2020-04554.
CAP acknowledges financial support from the Spanish Ministry of Science, Innovation and Universities (MICIU), under grants PID2023-149982NB-I00 and PID2023-146453NB-I00, and from the Agencia Estatal de Investigación (AEI) through the Severo Ochoa Centre of Excellence accreditation awarded to the Instituto de Astrofísica de Canarias, grant CEX2025-001609-S, funded by MICIU/AEI/10.13039/501100011033.
This research has made use of NASA's Astrophysics Data System Bibliographic Services and the \href{https://arxiv.org/}{arXiv} preprint server. 

This material is based upon work supported by the U.S. Department of Energy (DOE), Office of Science, Office of High-Energy Physics, under Contract No.~DE–AC02–05CH11231, and by the National Energy Research Scientific Computing Center, a DOE Office of Science User Facility under the same contract.
Additional support for DESI was provided by the U.S. National Science Foundation (NSF), Division of Astronomical Sciences under Contract No.~AST-0950945 to the NSF’s National Optical-Infrared Astronomy Research Laboratory; the Science and Technology Facilities Council of the United Kingdom; the Gordon and Betty Moore Foundation; the Heising-Simons Foundation; the French Alternative Energies and Atomic Energy Commission (CEA); the Secretariat of Science, Humanities, Technology and Innovation (SECIHTI) of Mexico; the Ministry of Science, Innovation and Universities of Spain (MICIU/AEI/10.13039/501100011033), and by the DESI Member Institutions: \url{https://www.desi.lbl.gov/collaborating-institutions}.
Any opinions, findings, and conclusions or recommendations expressed in this material are those of the author(s) and do not necessarily reflect the views of the U.~S.~National Science Foundation, the U.~S.~Department of Energy, or any of the listed funding agencies.

The authors are honored to be permitted to conduct scientific research on I'oligam Du'ag (Kitt Peak), a mountain with particular significance to the Tohono O’odham Nation.

This research made extensive use of \href{https://arxiv.org/}{arXiv.org} and the Science Explorer for bibliographic information.
The Science Explorer is funded by NASA under Cooperative Agreement 80NSSC21M00561.

This research has made use of the SIMBAD database, operated at CDS, Strasbourg, France \citep{Simbad}.
This research has made use of NASA’s Astrophysics Data System Bibliographic Services.

For the purpose of open access, the author has applied a Creative Commons Attribution (CC BY) licence to any Author Accepted Manuscript version arising from this submission.

{\it Software:} 
{\code{ArviZ} \citep{ArviZ2019}, 
\code{astropy} \citep{astropy,astropy:2018, astropy:2022}, 
\code{corner} \citep{corner2016},
\code{CmdStanPy}, 
\code{galpy} \citep{Bovy2015}, 
\textsc{jupyter} \citep{ipython, jupyter},
\code{matplotlib} \citep{matplotlib}, 
\code{numpy} \citep{numpy},
\code{pandas} \citep{pandas2022}, 
\code{scipy} \citep{2020SciPy-NMeth},
\code{seaborn} \citep{Waskom2021}, 
\code{Stan} \citep{Carpenter2017, stan-manual}
. 
}

Use of large language models (LLMs) in this work was governed by internal DESI Collaboration policy that emphasizes disclosure and ultimate responsibility of the scientist.
The authors used LLM-based tools to assist in the following tasks: cross-checking citations in ADS, code optimization, code debugging and reproducibility testing, rephrasing text for clarity, and proofreading manuscript text. 
The LLM used was Claude (Anthropic; Opus 5.5). 
All outputs were checked by the authors, and all scientific conclusions are the authors' own.


\bibliography{references,dr2papers} 
\bibliographystyle{aasjournal}



\appendix
\renewcommand\thefigure{\thesection\arabic{figure}}    
\setcounter{figure}{0}    

\renewcommand{\thetable}{\thesection\arabic{table}}
\setcounter{table}{0}    

\section{Bias correction factor across the AuriDESI realizations} 
\label{sec:fr_all}

\begin{figure*}[!hbt]
    \centering
    \includegraphics[width=\textwidth]{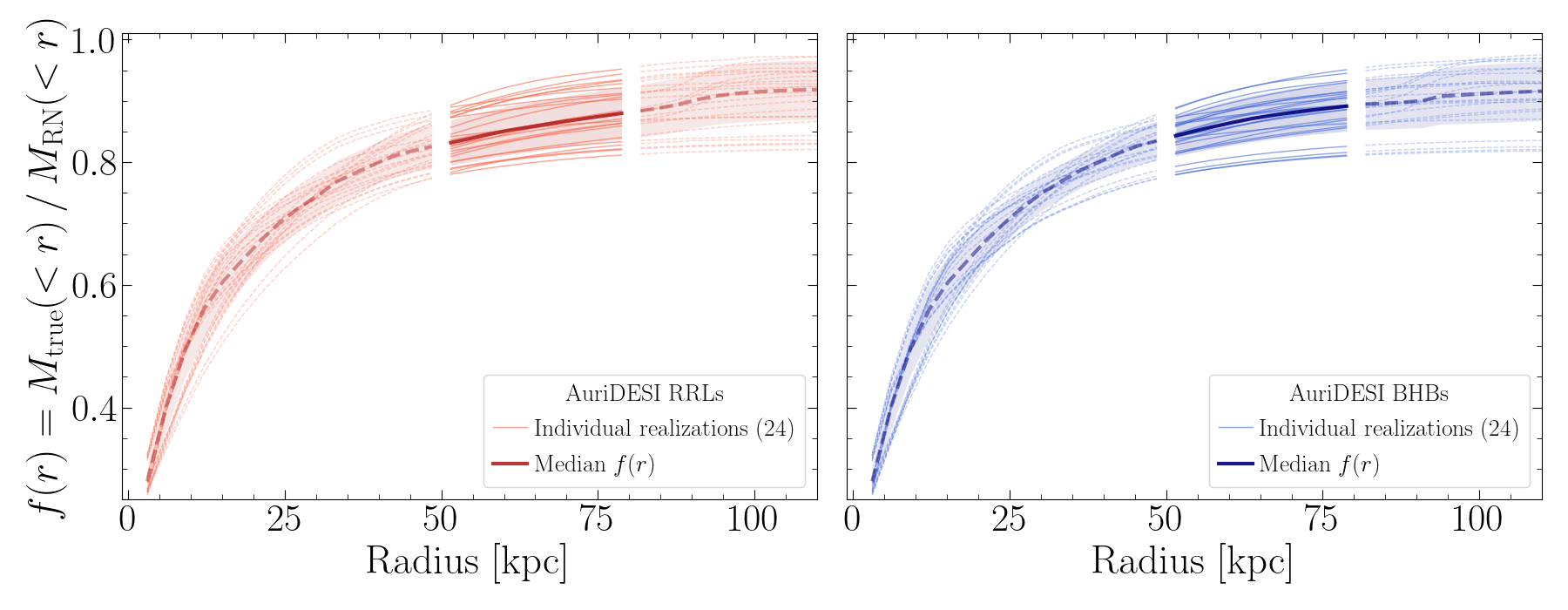}
    \caption{
    Correction factor $f(r) = M_{\rm true}(<r)/M_{\rm enc}^{\rm RN}(<r)$ for the 24 AuriDESI realizations for
    RRLs (\textit{left}) and BHBs (\textit{right}). 
    The values of $f(r)$ of the individual realizations are plotted with thin lines, while
    thick lines show the median correction factor $\langle f(r)\rangle$.  
    The $1\sigma$ scatter $\sigma_f(r)$ from these distributions is displayed as shaded regions. 
    Solid segments indicate the radial range covered by the tracers ($50$--$80$\,kpc), and dashed segments show the extrapolation of the fitted power-law profiles inwards and outwards.}
    \label{fig:fr_all24}
\end{figure*}

Figure~\ref{fig:fr_all24} shows the correction factor $f(r)$ defined in Section~\ref{sec:mockmass} for each of the 24 AuriDESI realizations, separately for RRLs and BHBs.
For both tracers, all 24 realizations overestimate the enclosed mass (i.e., $f<1$ at all radii in every mock halo and viewing angle). 
Within the radial range covered by the data ($50$--$80$\,kpc), the individual curves are smooth and nearly parallel, and the scatter between realizations reflects different amplitudes of the bias rather than different radial dependence.
Over this range, $\langle f(r)\rangle$ increases from $\sim\!0.8$ to $\sim\!0.9$ for both tracers (i.e., our uncorrected method overestimates the mock masses by $\sim\!20\%$ at 50\,kpc and $\sim\!13\%$ at 80\,kpc, with $\sigma_f\sim\!0.04$).

\setcounter{figure}{0}

\section{Posterior distributions of the GME parameters}

\begin{figure*}[!hbt]
    \centering    \includegraphics[width=0.49\textwidth]{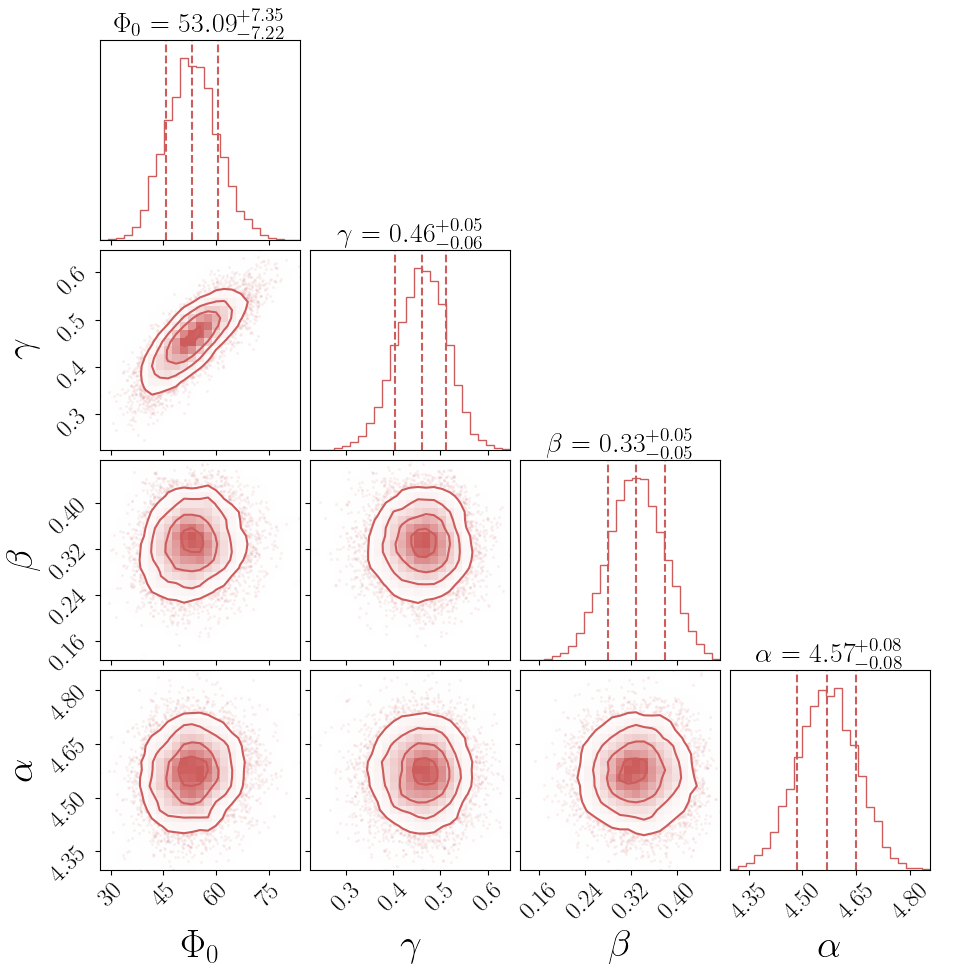}
   \centering    \includegraphics[width=0.49\textwidth]{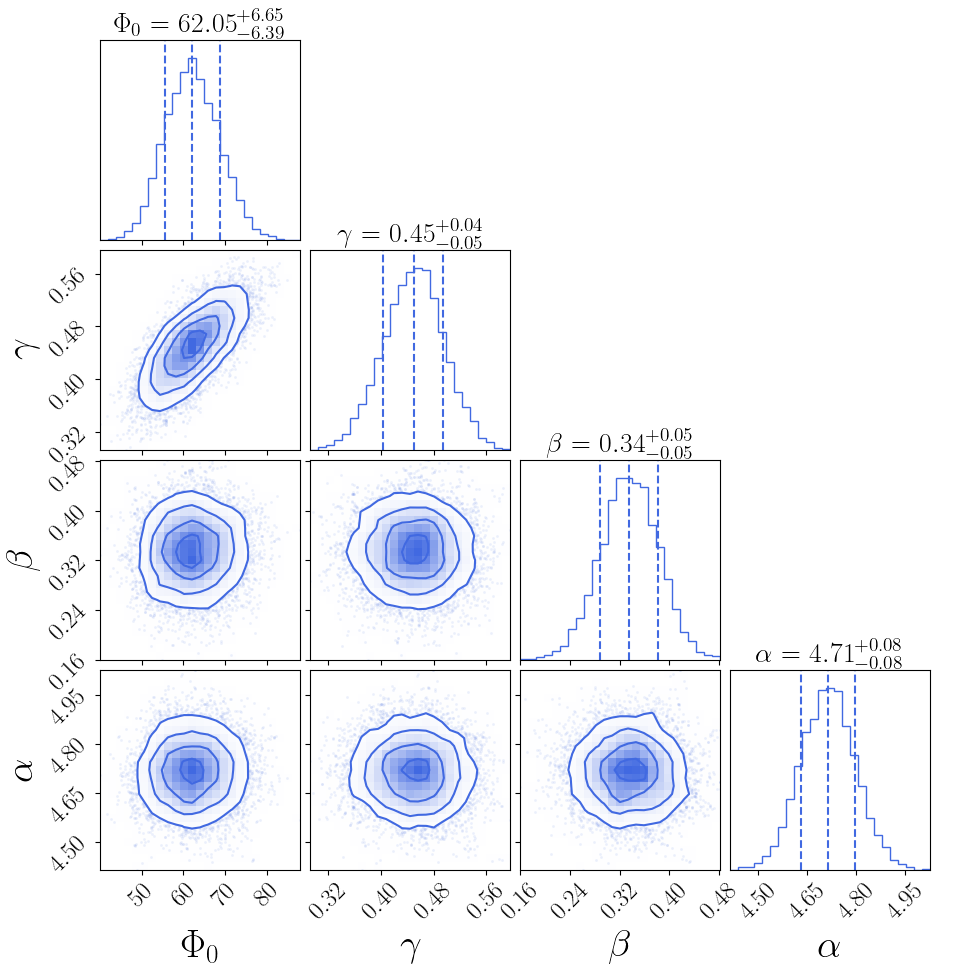}
    \caption{Marginal posterior distributions of the GME model parameters ($\Phi_0$, $\gamma$, $
    \beta$, and $\alpha$), shown for the DESI DR2 RRL (left) and BHB (right) samples.
    Vertical dashed lines mark the 16th, 50th, and 84th percentiles of each marginal distribution.
    }
    \label{fig:cornerplots}
\end{figure*}

Figure~\ref{fig:cornerplots} shows the joint and marginal posterior distributions of the four GME model parameters ($\Phi_0$, $\gamma$, $\alpha$, $\beta$) inferred from the DESI DR2 RRL and BHB samples. 
The posteriors are well-converged across all parameters. 
A notable covariance is present between $\Phi_0$ and $\gamma$ (most clearly seen in our RRL-based estimates), reflecting the degeneracy between the potential normalization and slope in the power-law potential (Equation~\ref{eq:enclosedmass}), as also reported by \citet{Shen2022} and \citetalias{Medina2025c}.

\setcounter{figure}{0}

\section{Extrapolating the GME constraint to $M_{200}$}
\label{sec:m200_computation}

\begin{figure}
    \centering    \includegraphics[width=1\columnwidth]{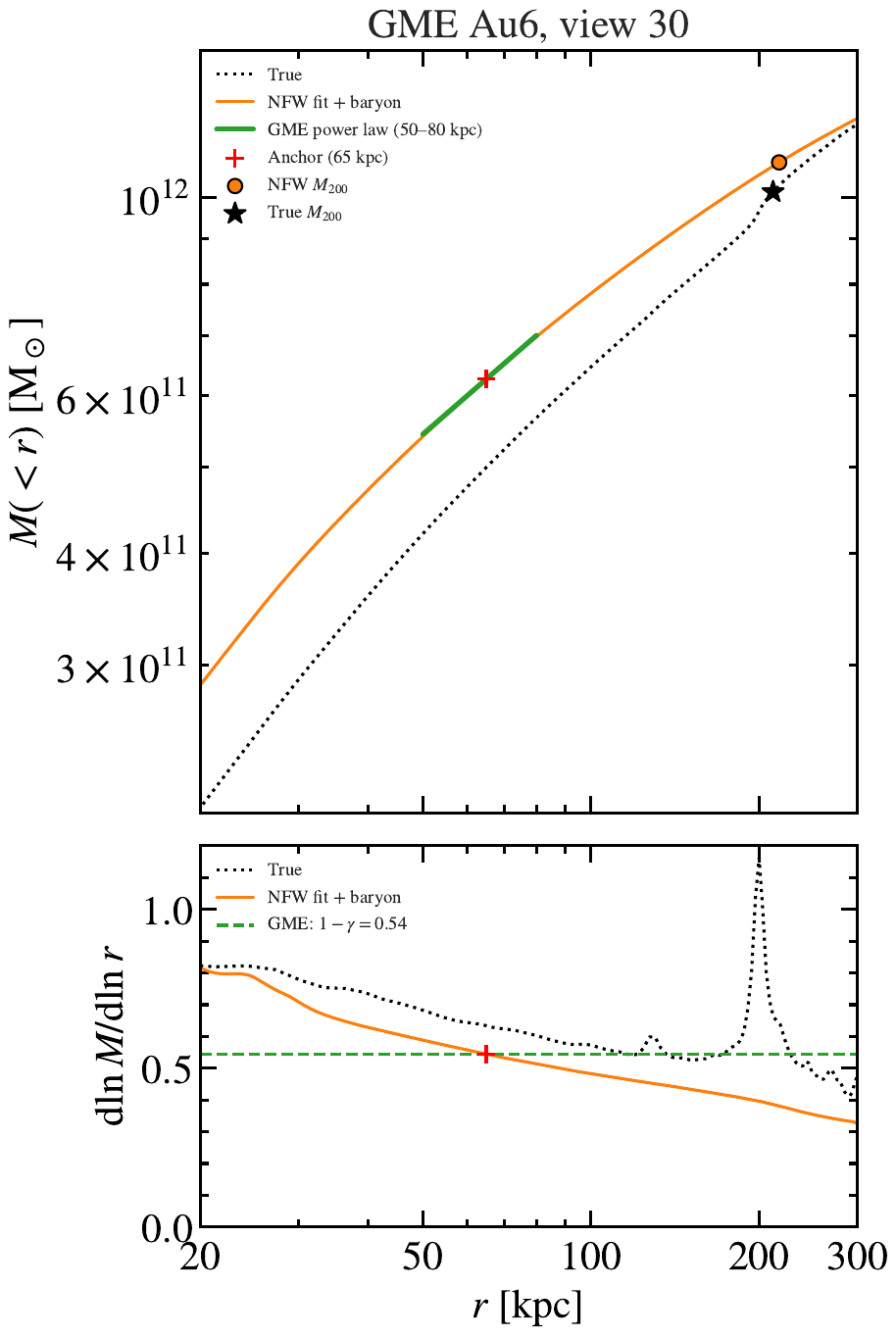}
    \caption{ The results of GME and the subsequent NFW fit for modelling the RRL sample of AuriDESI halo 6, $30\degr$.
    The top panel shows the total mass enclosed, the bottom the log slope with radius.
    The GME constraints of the total mass enclosed and slope values are shown in green,
    with the value at 65 kpc highlighted by the red cross scatter point.
    The orange line shows the NFW profile fit to these values, including a fixed baryonic profile.
    The true values from the simulation is shown in dotted black.
    See text for details.
    }
    \label{fig:AuM200Fit}
\end{figure}

Our GME analysis models the potential as a single power law within a radial bin, here $50$--$80\,\mathrm{kpc}$,
yielding a posterior chain over the amplitude $\Phi_0$ and slope $\gamma$.
The corresponding enclosed mass and slope is:
\begin{equation}
    M(<r) = \frac{\gamma \Phi_0}{G}\, r^{1-\gamma},
    \qquad
    \frac{\mathrm{d}\ln M}{\mathrm{d}\ln r} = 1 - \gamma .
\end{equation}
Outside this radial range the assumption of this power law becomes dubious, and the extrapolation unreliable.
In particular, a pure power law should not be extrapolated to the virial radius ($R_{200} \approx 220\,\mathrm{kpc}$),
which would systematically over-predict $M_{200}$ for realistic (NFW-like) halos,
whose mass profiles steepen beyond the fitted window.
Instead, we use the GME constraint as a local measurement of the enclosed mass and
its logarithmic slope at the anchor radius $r_0 = 65\,\mathrm{kpc}$, the centre of the fitted range.
To these estimates, we fit a physically motivated NFW halo and fixed baryon model.

For a fitted $(\Phi_0, \gamma)$ we proceed as follows.
First, we subtract a fixed baryonic contribution $M_{\rm b}(<r)$  from the GME total mass
and its radial derivative at $r_0$,
giving the dark-matter mass $M_{\rm dm}(<r_0)$ and slope $s_{\rm dm} \equiv \mathrm{d}\ln M_{\rm dm}/\mathrm{d}\ln r\,|_{r_0}$.
Second, we match these two quantities to an NFW halo.
The NFW enclosed-mass slope is,

\begin{align}
    \frac{\mathrm{d}\ln M_{\rm NFW}}{\mathrm{d}\ln r}
    &= \frac{x^2}{(1+x)^2 f(x)},
    \\
    f(x) &= \ln(1+x) - \frac{x}{1+x},
    \\
    x &= r/r_{\rm s},
\end{align}

\begin{equation}
    M_{\rm s}\, f(R_{200}/r_{\rm s}) + M_{\rm b}(<R_{200})
    = \frac{4}{3}\pi\, 200\, \rho_{\rm crit}\, R_{200}^3 ,
\end{equation}
defining $R_{200}$ from the combined dark-matter and baryonic mass,
and report the total mass $M_{200} = M(<R_{200})$.
We repeat this for our entire posterior chain of  $(\Phi_0, \gamma)$ to get a posterior distribution of $M_{200}$.
Samples for which the implied dark-matter slope falls outside the physical NFW range $0 < s_{\rm dm} < 2$ are discarded,
but in practice we find that this removes a negligible fraction of the chain.
Propagating every posterior sample through this mapping yields the full statistical distribution of $M_{200}$,
from which we quote the median and 16th--84th percentiles.

\begin{figure}
    \centering    \includegraphics[width=1\columnwidth]{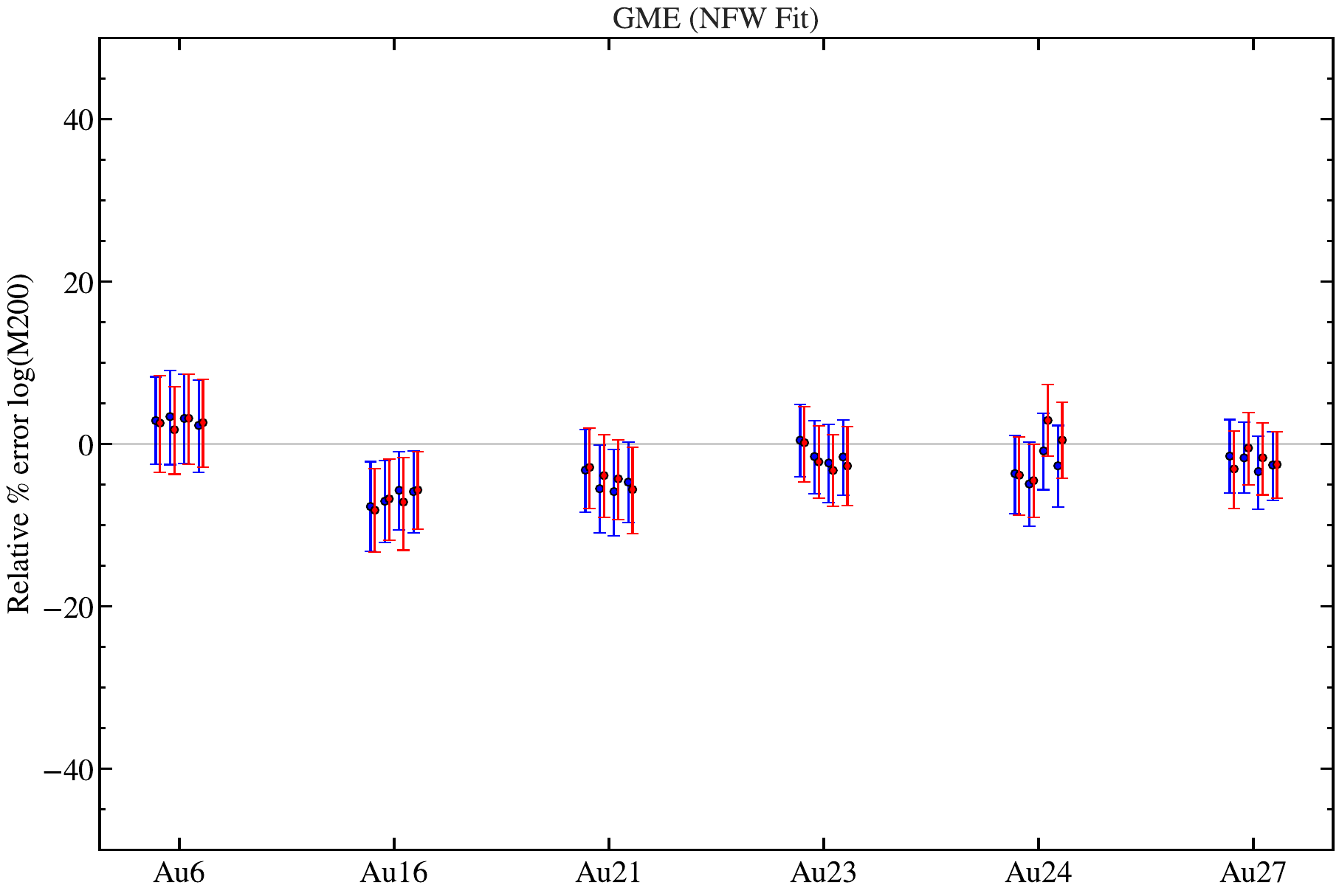}
    \caption{ The results of our M200 prediction across the AuriDESI sample.
    }
    \label{fig:AuM200Compare}
\end{figure}

\begin{figure}
    \centering    \includegraphics[width=1\columnwidth]{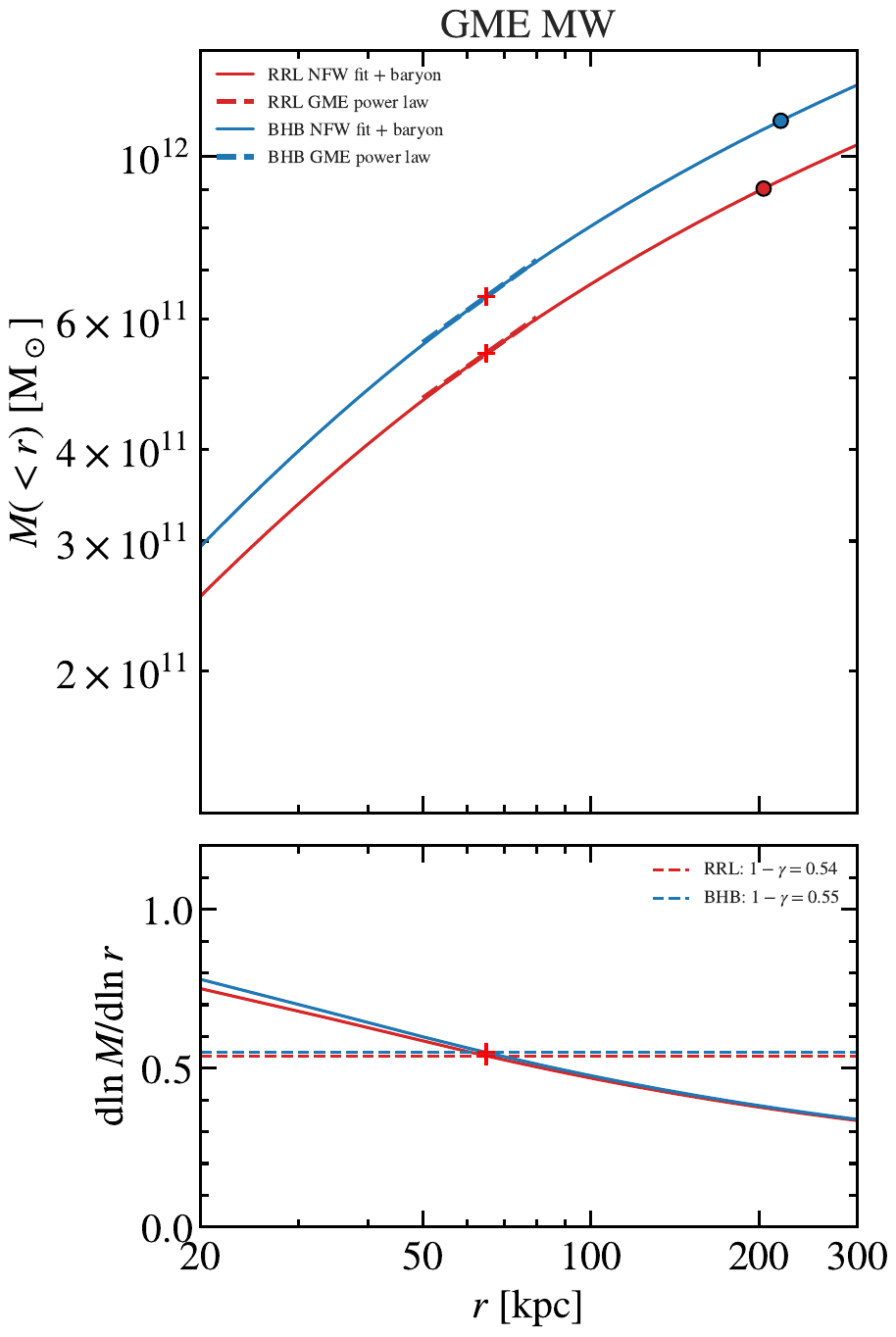}
    \caption{ Fitting the MW RRL and BHB samples.
    }
    \label{fig:MWM200Fit}
\end{figure}

\begin{figure}
    \centering    \includegraphics[width=1\columnwidth]{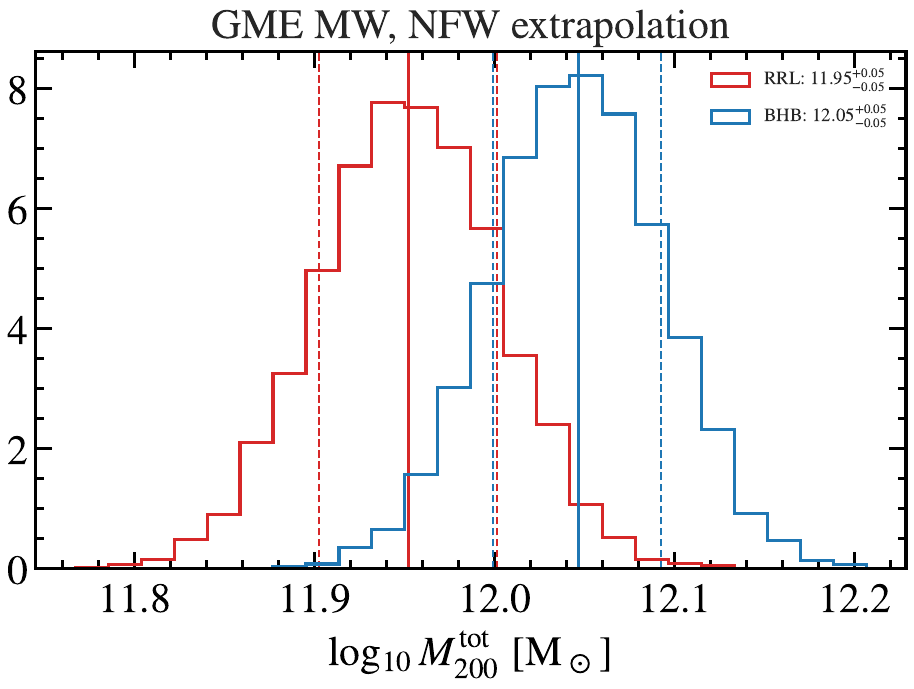}
    \caption{ The M200 results of our RRL and BHB sample in the Milky Way.
    }
    \label{fig:MWM200Results}
\end{figure}

Figure~\ref{fig:AuM200Fit} illustrates this procedure for the RRL sample of Au-6 (viewing angle of $30\degr$), and Figure~\ref{fig:AuM200Compare} compares the resulting $M_{200}$ estimates with the true values across the AuriDESI realizations.
The corresponding fits and $M_{200}$ posteriors for the DESI DR2 RRL and BHB samples are shown in Figures~\ref{fig:MWM200Fit} and \ref{fig:MWM200Results}.

\end{document}